\newcommand{\ltappeq}{\raisebox{-0.6ex}{$\,\stackrel
{\raisebox{-.2ex}{$\textstyle <$}}{\sim}\,$}}

\documentclass[12pt,preprint]{aastex}
\usepackage{psfig}

\slugcomment{to appear in the Dec 20 issue of ApJ}

\shorttitle{The Long-Period Cataclysmic Variable AKO~9 in 47 Tuc}
\shortauthors{Knigge et al.}

\begin{document}


\title{A Far-Ultraviolet Survey of 47 Tucanae. II. The Long-Period Cataclysmic Variable AKO~9
\footnote{Based on observations made with the
NASA/ESA Hubble Space Telescope, obtained at the Space Telescope
Science Institute, which is operated by the Association of
Universities for Research in Astronomy, Inc., under NASA contract NAS
5-26555. These observations are associated with proposals \#8219 and \#8267.}}


\author{Christian Knigge}
\affil{Department of Physics and Astronomy. University of Southampton, 
Southampton~SO17~1BJ,~UK}

\author{David R. Zurek, Michael M. Shara}
\affil{Department of Astrophysics, American Museum of Natural History, New York, 
NY 10024}

\author{Knox S. Long, Ronald L. Gilliland}
\affil{Space Telescope Science Institute, Baltimore, MD 21218}



\begin{abstract}

We present time-resolved, far-ultraviolet (FUV) spectroscopy and
photometry of the 1.1~day eclipsing binary system AKO~9 in the globular
cluster 47 Tucanae. The FUV spectrum of AKO~9 is blue and exhibits
prominent C~{\sc iv} and He~{\sc ii} emission lines. The spectrum
broadly resembles that of long-period, cataclysmic variables (CVs) in the 
galactic field. 

Combining our time-resolved FUV data with archival optical photometry
of 47~Tuc, we refine the orbital period of AKO 9 and define an accurate 
ephemeris for the system. We also place constraints on several other
system parameters, using a variety of observational constraints. We 
find that all of the empirical evidence is consistent with AKO~9 being
a long-period dwarf nova in which mass transfer is driven by the
nuclear expansion of a sub-giant donor star. We therefore conclude
that AKO~9 is the first spectroscopically confirmed CV in 47~Tuc.  

We also briefly consider AKO~9's likely formation and ultimate
evolution. Regarding the former, we find that the system was almost 
certainly formed dynamically, either via tidal capture or in a 3-body 
encounter. Regarding the latter, we show that AKO~9 will probably end
its CV phase by becoming a detached, double WD system or by exploding
in a Type~Ia supernova. 

\end{abstract}




\section{Introduction}
\label{intro}

The dense cores of globular clusters are expected to contain numerous
close binary systems (CBs) produced via dynamical 2- and 3-body
interactions, such as tidal capture (Fabian, Pringle \& Rees
1975). The resulting CB populations are key to our understanding of
globular cluster evolution: by giving up gravitational binding energy
to passing single stars, CBs provide the heat input that is needed to
reverse core collapse and drive their host clusters towards
evaporation (Hut et al. 1992).

Cataclysmic variables (CVs) are particularly important tracers of a
cluster's CB population. These systems are interacting binaries in
which a main sequence or slightly evolved secondary star loses mass
via Roche-lobe overflow to a white dwarf (WD) primary. CVs are important 
because they should be both numerous and relatively easy to find. For
example, di Stefano \& Rappaport (1994) predict the presence of well
over 100 active CVs formed by tidal capture in 47~Tuc, roughly 45 of
which should have accretion luminosities in excess of
$10^{33}$~erg~s$^{-1}$. However, early optical searches found only a
handful of CVs in all globular clusters combined (e.g. Shara et
al. 1996; Bailyn et al. 1996; Cool et al. 1998).

Recently, the hunt for CVs in globular clusters has become much more
successful. On the one hand, deep Chandra imaging of several clusters
has uncovered numerous X-ray sources, whose properties and (in some
cases) optical counterparts are consistent with those expected for CVs
(Grindlay et al. 2001a; Grindlay et al. 2001b; Pooley et al. 2002a;
Pooley et al. 2002b; Heinke et al. 2003). On the other hand, we have
recently begun a far-ultraviolet (FUV) survey of globular cluster
cores, which is very well-suited to the task of discovering CVs 
(Knigge et al. 2002; hereafter Paper I).

The first target of our survey was 47~Tuc, for which we obtained
time-resolved FUV imaging and slitless spectroscopy (the latter is
feasible because the move to the FUV reduces the crowding in the core
dramatically). Results from the imaging portion of these observations
were presented in Paper I and already demonstrated the potential of
the FUV waveband for CV searches: all 
previously suspected CVs in our field of view were found to have 
variable FUV counterparts, and several new CV candidates were
detected. Following on from this, the present paper contains first
results from the spectroscopic portion of our FUV survey of
47~Tuc. More specifically, we present time-resolved FUV spectroscopy
(supported by optical and FUV time-resolved photometry) of the
brightest FUV source in 47~Tuc, AKO~9. 

The nature of AKO~9 has been a long-standing puzzle. It is known to be an
eclipsing binary with an orbital period of about 1.1~days (Edmonds et 
al. 1996) and with a significant UV-excess (Auri\`{e}re, Koch-Miramond \&
Ortolani 1989; Paper I). Claims that it is associated with an 
X-ray source in the cluster have been made (Auri\`{e}re, Koch-Miramond
\& Ortolani 1989; Geffert, Auri\`{e}re \& Koch-Miramond 1997) and
refuted (Verbunt \&  Hasinger 1998), but this particular
controversy has now been settled by Chandra observations. These 
revealed that AKO~9 {\em is} an X-ray source, with $L_x \simeq 7.5 \times 
10^{30}$~erg~s$^{-1}$ (Grindlay et al. 2001a). Finally, and most
puzzlingly, Minniti et al.(1997) observed an "unusual brightening" of
AKO~9, during which the system's U-band flux increased by 2~mag in
less than two hours. If this brightening is interpreted as a genuine
increase in the system's luminosity, it poses serious problems for 
essentially all models for this system (e.g. Minniti et al 1997).

Here, we show that AKO~9 is a variable, blue emission line source,
whose FUV spectrum and other characteristics securely identify it as a
long-period, dwarf-nova-type CV. We also derive a precise orbital
ephemeris for the system, which reveals that the brightening described 
by Minniti et al. (1997) was simply an eclipse egress observed at a
time when AKO~9 was already in outburst.

\section{Observations}

\subsection{Far-Ultraviolet Photometry and Slitless Spectroscopy}
\label{fuvobs}

Our FUV survey of 47~Tuc consists of thirty orbits of STIS/HST
observations, comprised of six epochs of five orbits each (HST program 
GO-8279). The first epoch took place in 1999 September, and the gap 
between the first and second epochs was roughly 11 months; the  second
through sixth observations all occurred within a space of about 9~days.  
In each observing epoch, we carried out FUV imaging and slitless
spectroscopy. Typical exposure times in the FUV were 600~s for both
spectroscopy and imaging. Our program is therefore sensitive to
variability on time-scales ranging from minutes to months.

All of our FUV observations used the FUV-MAMA detectors and were taken
through the F25QTZ filter. This filter blocks geocoronal Ly$\alpha$,
OI 1304~\AA~and OI] 1356~\AA~emission which would 
otherwise produce a high  background across the detector in our
slitless spectroscopy. We used the G140L first-order grating for the
spectroscopic observations, yielding a dispersion of 0.584
\AA~pixel$^{-1}$ and a spectral resolution corresponding to roughly
1.2~\AA~(FWHM). The effective bandpass with this instrumental set-up
is roughly 1450~\AA~--~1800~\AA~for both imaging and spectroscopy. 

The 1024$\times$1024 pixel FUV-MAMA detector covers approximately
25\arcsec$\times$25\arcsec, at a spatial resolution of about
0.043\arcsec~(FWHM). Our field of view (FoV) was chosen to overlap
with archival HST observations of 47~Tuc and includes the cluster
center. The calibration of the FUV imaging observations has
already been described in Paper I, so we focus here on the extraction 
and calibration of the FUV spectra of AKO~9. 

Figure~1 shows an example of the slitless spectra we obtained. 
Each trail in this figure corresponds to the dispersed image of a FUV point 
source. The sharp cut-off at the left hand side of each trail is due
to the abrupt decrease in sensitivity around 1450~\AA,
where the quartz filter becomes opaque. AKO~9 is by far the brightest
source in this spectral image, and even a cursory look at the raw data
shows that there are two bright emission lines in its spectrum.\footnote{Another emission line source is also visible in the
spectral image, roughly 20\% below AKO~9. This source is the dwarf nova
V2 (Paresce \& de Marchi 1994; Shara et al. 1996). Its FUV properties
will be described and analyzed in a separate publication.}

Since AKO~9 is extremely bright and relatively isolated, its
spectrum can be extracted fairly straightforwardly from the spectral
images. In practice, this was done using the {\sc apextract} package
within {\sc iraf/noao/twodspec}. 
\footnote{{\sc iraf} (Image Reduction and Analysis Facility) is distributed
by the National Astronomy Optical Observatories, which are operated by
AURA, Inc, under cooperative agreement with the National Science
Foundation.}
More specifically, the tasks in this package were used to locate,
center, trace and extract the spectrum of AKO~9. For the last step, an
11-pixel extraction box was used; this is identical to the extraction
box used in the STIS pipeline reductions for long-slit, first-order
FUV spectroscopy. Background/sky subtraction was performed using a
simple, linear background estimate. 

The spectra were wavelength calibrated in two steps. First, we matched 
the sharp rise around 1450~\AA~in the raw (counts~s$^{-1}$) spectra to 
the shape of the inverse sensitivity function (i.e. the throughput) of
the STIS/FUV-MAMA/F25QTZ/G140L combination. In doing so, we assumed a
perfectly linear dispersion, which is a very good approximation for
the G140L grating. The inverse sensitivity functions we used were 
constructed using the {\sc calcspec} task within the
{\sc stsdas/hst\_calib/synphot} package. The latest version
of {\sc synphot} accounts for the sensitivity loss of the FUV-MAMA
detectors as a function of time, so we calculated a unique
throughput curve for each observing epoch. This first calibration step
produced a wavelength scale whose internal error was better than one  
pixel. However, when we analysed the wavelength calibrated spectra, we
found that all absorption and emission lines were blue-shifted by
roughly 5~\AA~($\simeq 9$~pixels). This offset suggests that the 
inverse sensitivity function for the F25QTZ filter in {\sc synphot} 
misrepresents the true throughput curve, at least in the 
vicinity of the turn-on of the filter near 1450~\AA. We corrected 
for this effect in the second step, by identifying 
two strong, narrow absorption lines in the spectrum of AKO~9 as
Si~{\sc II} 1527~\AA~and 1533~\AA~and shifting each spectrum to
optimally center these features at their expected locations. 
It is worth noting that the
1533~\AA~Si~{\sc II} line is associated with an excited state, and
therefore cannot be interstellar. By contrast, Si~{\sc II} 1527~\AA~
probably contains both interstellar and intrinsic components. We are
confident that the resulting absolute wavelength scale is good to
better than 1~\AA, as indicated by the good match between expected and
observed wavelengths of other spectral lines in the calibrated
data (see Figure~3 below). A key match in this context is that between the
measured (1671.5~\AA) and expected (1670.8~\AA) locations of the weak
interstellar Al~{\sc ii} absorption line. The radial velocity of the
cluster itself ($\simeq 19$~km~s$^{-1}$; Pryor \& Meylan 1993)
corresponds to about 0.1~\AA~and is therefore negligible for our
purposes.

%% The spectra were wavelength calibrated by identifying two strong,
%% narrow absorption lines in the spectrum of AKO~9 as Si~{\sc II}
%% 1527~\AA and 1533~\AA and shifting each spectrum to optimally center
%% these features  at their expected locations. It is worth noting that
%% the 1533~\AA~Si~{\sc II} line is associated with an excited state, and
%% therefore cannot be interstellar. By contrast, Si~{\sc II} 1527~\AA is
%% a commonly observed interstellar line, and probably contains
%% both interstellar and intrinsic components in our spectra. We are
%% confident that the resulting absolute wavelength scale is good to
%% better than 1~\AA, as indicated by the good match between expected and
%% observed wavelengths of other spectral lines in the calibrated
%% data (see Figure~3). A key match in this context is that between the
%% measured (1671.5~\AA) and expected (1670.8~\AA) locations of the
%% interstellar interstellar Al~{\sc ii} absorption line.

%\footnote{We also attempted to carry
%out the wavelength calibration with the {\sc wavecal} task within {\sc
%stsdas/calstis}. This requires a correction for the offset of AKO~9
%from the nominal reference position of the G140L grating. To estimate
%this offset, we used the the position of AKO~9 in the corresponding
%direct FUV images for each epoch. However, this method also did not
%solve the problem (and in fact made it somewhat worse). 

The flux calibration of the spectra was also performed in two
steps. First, we corrected for the wavelength-dependent ``slit 
losses'' caused by the finite size of our spectral extraction
box. Correction factors at several wavelengths were taken from the
appropriate photometric correction table that is provided by STScI as
a reference calibration file. Within our bandpass, the correction is
essentially linear in wavelength, which allowed us to estimate
correction factors for all wavelengths of interest. Second, each
spectrum was divided by the inverse sensitivity function for the
corresponding observing epoch. As a result of the
mismatch between the predicted and observed location of the turn-on in
the throughput function, our flux 
calibration is uncertain shortward of about 1500~\AA; we therefore 
ignore this region in the analysis that follows. 

Finally, we also calculated the integrated FUV magnitude that would be
produced by each calibrated spectrum in the STIS/FUV-MAMA/F25QTZ
photometric bandpass. This was done with the {\sc calcphot} task
within {\sc synphot} and allows us to combine the FUV spectroscopy and
photometry for the purpose of time-variability analysis. As in Paper
I, all FUV magnitudes given are on the STMAG system and correspond to
an infinite photometric aperture.

\subsection{Optical Photometry}
\label{optobs}

In order to complement our FUV observations, we extracted optical
light curves of AKO~9 from the HST observations of 47~Tuc described by
Gilliland et al. (2000). This data set consists of near-simultaneous,
time-resolved photometry in two broad-band filters: F555W ($\simeq$~V) 
and F814W ($\simeq$~I). The observations were taken over a time span
of 8.3 days, roughly 2 months before the first FUV observing
epoch, For details on the data reduction process, the reader is
referred to Gilliland et al. (2000).

For the purpose of calculating a long-term ephemeris for AKO~9, we 
also used the F336W ($\simeq$~U) light curve of AKO~9 from the data
of Gilliland et al. (1995). This data set was obtained in September
1993, i.e. six years before our first FUV observations. A full description
of these observations may be found in Gilliland et al. (1995) and
Edmonds et al. (1996).

\section{Results}
\label{results}

\subsection{FUV Light Curves}
\label{fuvlight}

Figure~2 shows the FUV light curve of AKO~9 over the course of our six
observing epochs, with filled (open) circles indicating magnitudes estimated
from the FUV spectroscopy (imaging). Note that the two types of
magnitude estimates agree very well, even though they were analysed
independently and required different calibration data and
reduction techniques.

The most obvious features in the FUV light curves are the two deep
($\simeq 3$~mag) primary eclipses seen in epochs 1 and 4. We will
use these in Section~\ref{ephem}~to derive an accurate orbital
ephemeris for AKO~9. For completeness, the orbital phases predicted by
this ephemeris for our FUV observing epochs are already shown in
Figure~2.

Figure~2 also reveals the presence of long-term changes in AKO~9's FUV
brightness. More specifically, the system was approximately 0.3~mag
brighter in Epoch~2 than in Epoch~1, which preceded it by roughly 11
months. In addition, AKO~9's FUV brightness declined by approximately 
0.6~mag over the 9 day interval between Epochs~2 and 6. There is also
evidence for stochastic variability on time-scales of tens of minutes to
hours, with typical (out-of-eclipse) rms values of 0.1~mag in any
given epoch. 
%Judging from Figure~2, this variability does not appear
%to be periodic, and this impression is confirmed by a power spectrum
%analysis of the data.

We emphasize that all types of variability highlighted in this
section are well in excess of instrumental uncertainties and must be
intrinsic to the source. This was verified explicitly by constructing
time series for a WD comparison star in the same way as for  
AKO~9. These reference light curves displayed significantly lower
levels of variability on all relevant time-scales, even though the
comparison star was 3~mag fainter than AKO~9 in the FUV.

\subsection{The Average FUV Spectrum}
\label{avespec}

Figure~3 shows the average FUV spectrum of AKO~9 away from
eclipse. This average was constructed by combining all FUV
spectra from all observing epochs, but excluding points affected by
the eclipses (see Figure~2). Note that, given the deep eclipses,
essentially all of the FUV flux must come from AKO~9's primary
(i.e. the eclipsed object). 

As expected from the raw, 2-D spectral image in Figure~1, the
calibrated 1-D spectrum contains extremely strong C~{\sc iv} and
He~{\sc ii} emission lines. Perhaps more surprisingly, the FUV
continuum turns out to be very blue. In order to quantify this
statement, we corrected the spectrum for the reddening towards
47~Tuc ($E_{B-V} = 0.032$; VandenBerg 2000) and then fit a power law 
model ($F_{\lambda} \propto \lambda^{\alpha}$) to the dereddened
continuum. 
\footnote{Here and below, we follow Paper~I in adopting the cluster
parameters suggested by VandenBerg (2000). However, our analysis is
not particularly sensitive to small changes in these parameters. Thus
our results would be much the same if we had instead adopted 
the parameters derived by Gratton et al. (2003), for example.}
The continuum windows used are indicated in the figure,
and the resulting fit yielded a spectral index of $\alpha=-2.74 \pm
0.11$. The (reddened) power law fit is also shown in Figure~3. 

Since our FUV spectra cover a relatively small wavelength range, it is
worth checking if the inferred continuum shape is consistent with the
observed optical characteristics of AKO~9. We did this by 
extrapolating the FUV spectral energy distribution into the optical
region and comparing the result to the observed optical magnitude. The
U-band is the best optical waveband to 
use for this purpose, both because it minimizes the extent of the
required extrapolation and because the primary still contributes a lot
of light in this bandpass (Edmonds et al. 1996). The extrapolation of
the power law fit to the U-band yields a predicted magnitude of $U
\simeq 18$ for the primary. This is comparable to AKO~9's
total out-of-eclipse magnitude in the normal state (Minniti et
al. 1997), and thus is certainly in the right ballpark, albeit
somewhat too bright. (The U-band eclipse depth is about 1 mag [Edmonds
et al. 1996], so the primary is expected to be roughly 0.6 mag fainter
at U than the combined system.)

\subsection{Long Time-Scale Changes in the FUV spectrum}
\label{slow}

Figure~4 shows the average FUV spectrum for each observing epoch,
along with the corresponding rms spectrum. Spectra affected by primary
eclipses were again excluded in constructing the averages, and the rms
spectrum was defined as 
\begin{equation}
F_{\lambda,rms} = 
\left[\frac{\sum_{i=1}^{N} \left( F_{\lambda,i} -
\left< F_{\lambda} \right> \right)^2}{N} \right]^{1/2},
\label{rms}
\end{equation}
with $N=6$~being the number of observing epochs.

Figure~4 clearly shows the overall brightening of the system between
Epochs~1 and 2, and its subsequent fading across Epochs~2 through 6. This
was already noted in Section~\ref{fuvlight} on the basis of the FUV
light curves. However, Figure~4 also reveals that the spectral
variability is not ``grey''. More specifically, the continuum shape of
the rms spectrum is significantly redder than the mean spectrum. A
power law fit to the dereddened rms spectrum, using the same continuum
windows as before (see Figure~3), yielded $\alpha=-2.0 \pm 0.2$. 

In order to quantify the continuum shape changes further, we again fit
power laws to the continua of the individual epoch averages. Figure~5
plots the resulting spectral index estimates against the corresponding
integrated fluxes ($F_{1500-1800}$), but reveals no correlation. This
is somewhat surprising. For example, if an increase in flux was due to
an increase in the characteristic temperature of the emitting region,
one might expect the spectrum to become bluer with increasing flux.  

The C~{\sc iv} and He~{\sc ii} emission lines are also variable, with
clear signatures from both lines being present in the rms
spectrum. However, the bottom panel of Figure~4 reveals that the
{\em fractional} variability of He~{\sc ii} is actually
indistinguishable from that of the adjacent continuum, and that the
fractional variability of C~{\sc iv} is actually {\em less} than that
of the adjacent continuum. This may be a sign of saturation in C~{\sc
iv}.

\subsection{The FUV Spectrum during Eclipse}
\label{phasespec}

With two eclipses in our FUV data (Figure~2), we can track the change
in the FUV spectrum over the course of an eclipse. Figure~6 shows the
average out-of-eclipse, ingress, and mid-eclipse spectra
derived from the FUV data. The out-of-eclipse spectrum is the
same as that shown in Figure~3. The ingress spectrum was defined as the
mean of those spectra that produced FUV magnitudes in the range $m_{FUV} =
16.5 -- 18.0$~(see Figure~2). Finally, the mid-eclipse spectrum was
defined as the mean of those spectra yielding $m_{FUV} > 18.0$.

The key point to note from Figure~6 is that the two strong emission
lines are much less deeply eclipsed than FUV continuum. This implies
that these lines must be formed in a region whose size is comparable
to that of the secondary star. This behaviour is quite common among
field CVs and is generally ascribed to line formation taking place in
a vertically extended region, such as an accretion disk wind
(e.g. Knigge \& Drew 1997). 

It is also worth noting that both emission lines show evidence of
being formed in a rotating medium. This is illustrated in Figure~7,
which compares the out-of-eclipse and ingress profile shapes of both 
C~{\sc iv}~and He~{\sc ii}~lines. In both cases, the blue line wing of
the ingress profile is suppressed relative to the red wing. This
behaviour is expected during the eclipse of a line forming region that
is rotating in the same sense as the binary system. Similar 
rotational disturbances are often seen in the disk-formed
optical lines of field CVs (e.g. Baptista, Haswell \&  Thomas 2002;
Thorstensen 2000), but have also been observed in the C~{\sc iv} line
of at least one such system (UX~UMa; Baptista et al. 1995; Knigge \&
Drew 1997).

\subsection{Optical Light Curves}
\label{optlight}

The V- and I-band light curves of AKO~9 obtained from the Gilliland et 
al. (2000) data are shown in Figure~8. A strong signal on half the
orbital period (roughly 0.55 days) is easily seen in both
bandpasses. We therefore obtained a first estimate of the orbital
period by directly fitting sine waves to the light curves. This yielded
$P_{orb} = 1.1080 \pm 0.0013$~days, which is consistent with previous
estimates (e.g. Edmonds et al. 1996). 

Figure~9 shows the result of folding the data onto this period
estimate. As expected, a double-humped modulation is present in both V
and I, but a weak primary eclipse is also seen in both filters. The
dominant signal on $P_{orb}/2$ is suggestive of an ellipsoidal
modulation due to the distorted shape of a Roche-lobe-filling
secondary. The main cause of ellipsoidal variations is the changing projected
area such a star presents to a distant observer. Since the 
Roche lobe is elongated along the binary separation vector, maximum
projected area (and hence maximum 
brightness) is observed twice per orbit, at the quadrature phases
$\phi = 0.25$ and $\phi = 0.75$ (measured relative to the inferior
conjunction of the secondary; in an eclipsing system like AKO~9, this
coincides with the primary eclipse). The light curves in Figure~9 are
entirely consistent with this picture and thus support the idea that
AKO~9 is an {\em interacting} binary system.

In order to determine the intrinsic amplitude of the ellipsoidal
modulations, we fit the phase-folded light curves with a simple
analytical model consisting of three components: (1) a constant
(representing the average light from the secondary); (2) a sinusoid
with period 
$P_{orb}/2$ (representing the ellipsoidal variations); (3) a constant
with a parabolic eclipse (representing the primary). The best model
fits are indicated in Figure~9. The model is clearly simplistic, but 
its key parameters are robustly determined by the data. We find
that the secondary contributes approximately 90\% of the flux at V and
94\% of that at I. The intrinsic peak-to-peak amplitudes of the
ellipsoidal variations (measured in magnitudes and corrected for the
dilution of 
the signal by light from the primary) are $\Delta m_V = 0.24 \pm 0.03$ and
$\Delta m_I = 0.16 \pm 0.03$. These measurements will be used in
Section~\ref{qi} to constrain the mass ratio and inclination of AKO~9.

\subsection{Orbital Period and Ephemeris}
\label{ephem}

The orbital period estimate derived in the previous section can be
improved considerably by using the optical and FUV data in
concert. More specifically, we first used parabolic fits to estimate
the times of mid-eclipse for the two eclipses seen in the FUV
data. The resulting eclipse timings are listed in Table~1. The optical
data is too sparse to warrant similar timing measurements for
individual eclipses. Instead, we used a parabolic fit to determine the
mid-eclipse {\em phase} in the folded V-band light curve (Figure~9); the
I-band light curve was not used for this purpose since the eclipse
is poorly sampled by the folded light curve. We then
combined this mid-eclipse phase with the existing orbital period
estimate to yield a single, accurate mid-eclipse timing to represent
the collective V-band data. This ``reference eclipse'' was chosen to
lie near the middle of the V-band time series. 

Armed with these three mid-eclipse timings, we used the orbital period
estimate obtained from the optical data (Section~\ref{optlight}) to
calculate the corresponding cycle counts. The
cycle count separating the two FUV eclipses turned out to be
ambiguous at this stage, but that separating the V-band eclipse and the first FUV
one was already secure. This cycle count was therefore replaced with an
integer and used to refine the orbital period. With this refined period,
the ambiguity of the second cycle count was also removed, and integer
cycle counts could be assigned to all three eclipse timings. A
weighted least squares fit to the timings (with
cycle count as the independent variable) was then used to improve the
orbital period further. As a further refinement, we recalculated the
mid-eclipse timing of the V-band reference eclipse, i.e. the latest
period estimate period was used to re-fold the data and to convert the 
measured mid-eclipse phase to the corresponding mid-eclipse
timing. The resulting timing estimate is also given in Table~1. A new 
orbital period estimate was then obtained from another weighted
least squares fit to all three timings. The corresponding ephemeris is
given by 
\begin{equation}
T_{mid-eclipse} (BJD) = 2451546.7767(13) \; + \; 1.1090989(74) \; E
\label{eq_ephem1}
\end{equation}
where the numbers in parentheses are the 1-$\sigma$~joint errors on 
the last two digits of the two fit parameters. The zero point of this
ephemeris has been chosen to lie near the weighted average of the
eclipse timings; this removes the correlation between zero point and
period errors. Barycentric corrections were calculated using John
Thorstensen's {\sc skycalc}~package.

At this stage, we have not yet used the 1993 U-band light curve of
AKO~9 obtained by Gilliland et al. (1995), which also covers one
complete eclipse (see their Figure~12c and Figure 6 in Edmonds et 
al. 1996). We therefore carried out another parabolic fit to estimate
the time of mid-eclipse in this data set (see Table~1). Applying 
Equation~\ref{eq_ephem1} and solving for the corresponding cycle count 
yielded $E = -2083.056 \pm 0.014$. Clearly, there is no 
cycle count ambiguity, but the ephemeris prediction is inconsistent
with the observed timing at about the 4-$\sigma$~level. If we simply
add the 1993 U-band timing to the three more recent timings and
recalculate the ephemeris, we obtain 
\begin{equation}
T_{mid-eclipse} (BJD) =  2451441.4095(72) \; + \; 1.109125(15)  \; E.
\label{eq_ephem2}
\end{equation}
However, this linear fit to the timing data is quite poor, with
$\chi^2 = 121.1$~(for four data points and two free parameters). For
comparison, the fit described by Equation~\ref{eq_ephem1} returned 
$\chi^2 = 3.6$~(for 3 data points and two free parameters). A
quadratic fit to the data is formally better, but implies an extremely
short time-scale for period change (on the order of $10^5$~yrs). We
refrain from quoting this or any other non-linear ephemeris fits
because we still consider our data set too sparse for such
applications. However, we will show in Section~\ref{brightening} that
another eclipse timing from data obtained in 1995 is also not very
well described by Equation~\ref{eq_ephem1} (with the sense of the
discrepancy being the same as for the 1993 timing). Thus the gathering
of additional eclipse timings is strongly encouraged. In the meantime,
we recommend Equation~\ref{eq_ephem1} for the prediction and analysis
of future eclipse timings.

\subsection{System Parameters}
\label{syspars}

In this section, we provide estimates for the system parameters of the
AKO~9 binary system. We note from the outset that most of these
estimates will rely explicitly or implicitly on AKO~9 being a
semi-detached system. We regard this assumption as secure (see also
Section~\ref{dn}). A summary of the key constraints obtained in this
section is given in Table~2.

\subsubsection{The Parameters and Evolutionary State of the Secondary}
\label{secpars}

The volume-averaged radius of the Roche-lobe-filling star in a
semi-detached binary system depends only on the binary separation, $a$, and
the mass ratio, $q = M_2/M_1$ (e.g. Frank, King \& Raine
2002). Eggleton (1983) has provided a useful fitting formula for this
dependence, which is valid for all mass ratios:
\begin{equation}
\frac{R_2}{a} = \frac{0.49 q^{2/3}}{0.6 q^{2/3} + \ln{(1 + q^{1/3})}}.
\label{egg}
\end{equation}
This equation can be combined with Kepler's third law to solve for the
mean density of the secondary. The numerical result can be
conveniently expressed analytically as  
\begin{equation}
\left<\rho\right> = \frac{M_2}{\frac{4}{3}\pi R_2^3} = 0.16 \; f(q) \; 
\left(\frac{P_{orb}}{1.109~\rm{day}}\right)^{-2} {\rm ~g~cm^{-3}}, 
\label{dense}
\end{equation}
where $M_2$ and $R_2$ are the mass and radius of the secondary, and
$P_{orb}$ has been scaled to AKO~9's orbital period. The function
$f(q)$ has a maximum of unity at $q = 0.21$ but varies only mildly
over the interesting range of mass ratios (e.g. f(0.02) = 0.92;
f(2.00) = 0.75). We can therefore take $\left<\rho\right> \simeq 0.15~{\rm
g~cm^{-3}}$ as an estimate of the secondary's average density. 
At 47~Tuc's metallicity ([Fe/H] = -0.83; VandenBerg 2000), this
density corresponds to that of a roughly 40~$M_{\odot}$ zero-age main
sequence (MS) star (as determined from the mass-radius relationship
given by Tout et al. 1996). Obviously, all such stars have evolved
off 47~Tuc's MS long ago, since the turn-off occurs at about $M_{to}
\simeq 0.9 M_{\odot}$ (VandenBerg 2000). Thus AKO~9's secondary must
be an evolved object. 

We can use the optical data to shed additional light on the nature of
the secondary. In Section~\ref{optlight}, we estimated that the secondary
contributes roughly 90\% (94\%) of the V-band (I-band)
light. Edmonds et al.(2003) cite $V = 17.40$ and $V-I = 0.96$ for
AKO~9 based on the same data we have used here. Correcting this for
the contribution of the primary, we find $V_2 \simeq 17.5$ and $(V -
I)_2 \simeq 1.0$ for the secondary alone. This puts the system
close to, but slightly below and to the red of, 47~Tuc's sub-giant
branch in the corresponding color-magnitude diagram (CMD; Albrow et
al. 2001; Edmonds et al. 2003).  

Albrow et al. (2001) have suggested that this CMD location is
consistent with that expected for a mass-losing sub-giant
secondary. We therefore checked if this idea is in line with the
density constraint derived above. VandenBerg (2000) has shown that
$\alpha$-enhanced stellar models on an 11.5~Gyr isochrone provide a
good fit to the composite cluster CMD reported by Hesser et
al.(1987). We therefore calculated
the densities of stars on this isochrone by interpolating on
VandenBerg's (2000) models, which he kindly provided for us. We 
found that $\left<\rho\right> =  0.15~{\rm g~cm^{-3}}$ does indeed
correspond to a star near the end of 47~Tuc's sub-giant branch, with
parameters $M = 0.9~M_{\odot}$, $R = 2.0~R_{\odot}$, $\log{g} = 3.8$
and $T_{eff} = 5300$~K. 

However, as expected from the CMD location, an ordinary sub-giant is
somewhat too bright and too blue to match the V-band magnitude and 
V-I color of AKO~9's secondary. We therefore used the synthetic
photometry package {\sc synphot} within {\sc iraf/stsdas}
to search for a stellar model that does match these observations. We
found that a model with $T_{eff} = 4900$~K and $R = 2.1~R_{\odot}$ succeeds 
in this respect (the surface gravity is essentially unconstrained by the
photometry). With $M = 
0.9~M_{\odot}$, the density of such a star is $\left<\rho\right> =
0.14~{\rm g~cm^{-3}}$ and thus still satisfies the density
constraint. The gravity is then also effectively unchanged from the 
standard sub-giant model, i.e. $\log{g} = 3.8$. We regard these
parameters as best-bet estimates for the secondary.

\subsubsection{Mass Ratio and Inclination}
\label{qi}

There are two ways in which our data can be used to constrain the mass
ratio and inclination of AKO~9. The first is based on the fact that
the system is both semi-detached and eclipsing. The full width at half
depth ($\Delta_{\phi}$) of the primary eclipse therefore 
depends only on $q = M_2/M_1$~and $i$ (e.g. Dhillon, Marsh \& Jones
1991). Thus a measurement of this parameter admits only a 
single family of solutions in the (q,i) plane (see Figure~2 in Horne 1985). 
In the case of AKO~9, we obtained $\Delta_{\phi} = 0.090 \pm 0.002$ from 
the optical and FUV eclipses. We then used a method similar to Chanan, 
Middleditch \& Nelson (1976) to calculate the corresponding family of
($q,i$) solutions. The result is shown in Figure~10. We also plot in
this figure the lower limit on the inclination that is implied by
the sheer existence of eclipses. This is given by $i >
\cos^{-1}{R_2/a}$, with $R_2/a$ as in Equation~\ref{egg}.

The second way to constrain $q$ and $i$ relies on the amplitude of
ellipsoidal variations. The intrinsic 
\footnote{By `intrinsic'', we mean ``corrected for any dilution
of the signal due to light sources other than the secondary''.} 
amplitude of the ellipsoidal modulation produced by a
Roche-lobe-filling star in a given passband depends on four
parameters: (1) the mass ratio; (2) the inclination; 
(3) the strength of the limb-darkening effect; and (4) the strength of 
the gravity-darkening effect. 

The strength of limb-darkening is usually parameterized by the linear
limb-darkening coefficient $u_l$. This is defined by $I(\mu)/(I(\mu = 1) = 1
- u_l(1 - \mu)$, where $I$ is the specific intensity and $\mu$ is the 
cosine of the angle between the line of sight and the
local normal to the stellar surface. We estimated appropriate values 
of $u_l$ by interpolating on the tables of monochromatic limb-darkening 
coefficients provided by van Hamme (1993).
In so doing, we assumed the secondary star parameters listed in Table 2
and took the pivot wavelengths of the F555W ($\simeq V$) and F814W
($\simeq I$) filters to represent their respective bandpasses. The
resulting estimates are $u_{l,v} = 0.68$ and $u_{l,i} = 0.49$.

The strength of gravity darkening can be measured by defining a
(wavelength-dependent) gravity darkening coefficient $u_g =
d\ln{F_\lambda } / d \ln{g}$. Note that this coefficient is closely
related to, but different from, the gravity darkening exponent $\beta 
= d \ln{T} / d \ln{g}$~and the ``bolometric'' (wavelength-integrated)
gravity darkening coefficient $u_{g,bol} = d\ln{F_{bol}} / d \ln{g} =
4 \beta$, both of which are also often quoted in the literature. Lucy
(1967) has shown that $\beta = 0.08$~for stars with convective
envelopes. We expect the sub-giant secondary of AKO~9 to fall into
this category, so we adopted this value of $\beta$~and used Equation~10
of Morris (1985) to convert this to $u_g$~values corresponding to the
pivot wavelengths of the F555W (V) and F814W (I) bandpasses. In the process, 
we again adopted $T_{eff} = 4900~K$~for the secondary. The resulting  
coefficients are $u_{g,v} = 0.43$~and $u_{g,i} = 0.30$.  
  
With these parameters fixed, the ellipsoidal amplitudes measured in
Section~\ref{optlight} can be converted into constraints on $q$ and
$i$. This final step was carried out with the aid of the relevant 
tables in Bochkarev, Karitskaya \& Shakura (1979), who provide
numerically calculated ellipsoidal amplitudes for a wide range of
parameters. However, since their tables are quite coarsely spaced, we
used the analytical approximation due to Russell (1945)
%\footnote{See also Morris (1985) for an apparently independent
%derivation of an almost identical approximation.} 
to fix the functional form of the amplitude with respect to~$q$
and~$i$. The Russell approximation can be written as (e.g. Drew,
Jones, Woods 1993) 
\begin{equation}
\Delta m = \frac{3}{2}\frac{f^3}{q}\left(\frac{R_2}{a}\right)^3
\sin{i}^2 (1 + u_g) \frac{15 + u_l}{15 - 5u_l}, 
\label{russell}
\end{equation}
where $f$ measures the fractional size of the secondary relative to
the Roche lobe; for a Roche-lobe-filling secondary, $f = 1$. We
therefore rescaled this equation to 
best match the numerical results of Bochkarev et al. (1979) before 
using it to calculate the constraints on $q$~and $i$~implied by the
observed ellipsoidal amplitudes. These constraints are shown as shaded
areas in Figure~10.

We note in passing that the rescaling of the Russell equation is quite
important: in the region of the ($q,i$) plane relevant to AKO~9, the 
Russell approximation typically underpredicts the true ellipsoidal
amplitude by a 
factor of about 1.6. This is not particularly surprising, since the 
approximation becomes increasingly inaccurate as the filling factor
$f$ approaches unity (see Morris 1985; Fig. 1). Despite this, the
Russell equation 
has occasionally been used in the literature to directly constrain
$q$~and $i$~in semi-detached systems. Such estimates should be
viewed with some suspicion and are probably biased too low (high)
in mass ratio (inclination). 

Combining the eclipse and ellipsoidal constraints, we arrive at our
final constraints on mass ratio and inclination: $0.6 < q < 2.6$ and
$68^o < i < 80^o$. Note that these are still, in essence,
1-$\sigma$ error ranges, although we have been conservative in
considering as valid all solutions that satisfy either of
the two (V and I) 1-$\sigma$ ellipsoidal constraints. It should also
be kept in mind that $q$~and $i$ cannot be varied independently, but
are constrained to lie along the line determined by the eclipse
width. The best-bet ($q$, $i$) pair (which lies at the intersection of
the V and  I constraints) is $q = 1.2$ and $i = 74^o$.

\subsection{FUV Flux and Luminosity}
\label{lum}

AKO~9 is the brightest FUV source in the core of 47~Tuc by almost
2~magnitudes. Its integrated FUV flux between 1500~\AA~and 1800~\AA~was
measured from the (dereddened) average, out-of-eclipse FUV spectrum
and is $F_{1500-1800} = 9.6 \times
10^{-13}$~erg~s$^{-1}$~cm$^{-2}$. If the power law fit described in 
Section~\ref{avespec} is used to extrapolate the spectrum to the
``full'' FUV range of 1000~\AA~--~2000~\AA, the integrated flux
increases to $F_{1000-2000} = 4.4 \times 10^{-12}$~erg~s$^{-1}$~cm$^{-2}$.

At the distance of 47~Tuc (4.5~kpc; VandenBerg 2000) these fluxes
correspond to FUV luminosities of $L_{iso(1500-1800)} = 2.3 \times
10^{33}$~erg~s$^{-1}$ and $L_{iso(1000-2000)} = 1.1 \times
10^{34}$~erg~s$^{-1}$ for an isotropically emitting source. However,
if AKO~9 is a CV (as we will argue in Section~\ref{dn}), the
assumption of isotropy may be quite poor. As an alternative, we
consider the possibility that the FUV light is produced by a
geometrically thin, 
optically thick accretion disk. It is fairly straightforward to
derive the equivalent of the usual $4 \pi d^2$~factor for the 
case of an emitting source with a disk geometry. The result, expressed
relative to the luminosity of an isotropic source producing the same
observed flux, is 
\begin{equation}
L_{disk} = \frac{L_{iso}}{2 * \eta (i) \cos{i}}. 
\label{corr}
\end{equation}
Here, $\eta (i)$ is the appropriate limb-darkening law, which we take
to be of the form $\eta (i) = 0.5 (1.0 + 1.5 \cos{i})$ (e.g. Knigge,
Woods \& Drew 1995). 
\footnote{Note that this limb-darkening law is defined relative to the
angle-averaged intensity, rather than to the face-on intensity
(cf. the linear limb-darkening law used to define $u_l$ in
Section~\ref{qi}).} On substituting the inclination constraints
derived in the previous section into Equation~\ref{corr}, we find that
the luminosity estimates should be increased by a factor 1.7 -- 4.0 if
AKO~9's FUV emission is produced by an optically thick disk.

\section{Discussion}
\label{discussion}

\subsection{Classification of AKO~9 as a Dwarf Nova}
\label{dn}

We begin this section by collecting the key observational properties of AKO~9: 
\begin{enumerate}
\item[(i)] it is a 1.1~day eclipsing binary (Edmonds et al. 1996);
\item[(ii)] it is the brightest FUV source in 47~Tuc (Paper I), with $L_{1500-1800} > 2.3 \times 10^{33}$~erg~s$^{-1}$ (Section~\ref{lum});
\item[(iii)] its FUV spectrum displays strong C~{\sc iv} and He~{\sc
ii} emission lines (Section~\ref{avespec});
\item[(iv)] the emission lines display a classic rotational
disturbance (Section~\ref{phasespec});
\item[(v)] its FUV continuum decreases steeply towards longer
wavelengths (Section~\ref{avespec});
\item[(vi)] its FUV flux is variable on both short (tens of minutes) and
long (days/months) time-scales (Sections~\ref{fuvlight});
\item[(vii)] it is a soft X-ray source with $L_x \simeq 7.5 \times
10^{30}$~erg~s$^{-1}$ (Grindlay et al. 2001a);
\item[(viii)] its optical colors are broadly consistent with those of a
cluster sub-giant (Albrow et al. 2001; Edmonds et al. 2003;
Section~\ref{secpars}); 
\item[(ix)] its optical light curves display double-humped ellipsoidal
variations (Section~\ref{optlight});
\item[(x)] it has on two occasions been caught in a high state, during
which it was approximately 2~mag brighter (in U) than in its normal
state (Minniti et al. 1997). 
\end{enumerate}

All of these properties point towards AKO~9 being a cataclysmic
variable, in which a WD primary is accreting from a Roche-lobe-filling
secondary. In particular, the eruptions described by Minniti et
al. (1997) strongly suggest that AKO~9 is a dwarf-nova-type CV. No
other class of binary system matches the
observational constraints nearly as well. We note, in particular, that
AKO~9 is unlikely to be a transient low-mass X-ray binary system
(LMXB) with  either a neutron star (NS) or a black hole (BH)
primary. This is because AKO~9's X-ray luminosity is much lower than
that of quiescent NS LMXBs (e.g. Campana et al. 1998), and its X-ray
colors are redder than expected for a quiescent BH LMXB (Kong et al. 2002). 

AKO~9's long orbital period and evolved secondary make it a rather
unusual CV. It is therefore worth comparing it more directly to its 
closest cousins in the galactic field. 
Such a comparison is shown in Figure~11, where the FUV spectrum of
AKO~9 is shown alongside those of two field CVs, BV~Cen
and GK~Per. The orbital periods of these systems bracket
that of AKO~9, with $P_{orb} = 0.61$~d for BV~Cen (Gilliland 1982) and
$P_{orb} = 2.00$~d for GK ~Per (Crampton, Fisher \& Cowley Fisher 1986). With
such long periods, both systems must also have evolved secondaries.

The field CV spectra are {\sc newsips} extractions of observations
obtained with the {\em International Ultraviolet Explorer} (IUE) 
and were taken from the IUE archive. The spectrum of 
BV Cen is IUE exposure SWP26623; that of GK Per is a weighted average
of exposures SWP15098, SWP15331 and SWP18226. Both BV Cen and GK Per
are dwarf novae, but all exposures used here were taken during 
quiescent periods to ensure a fair comparison with the spectrum of
AKO~9. The spectra of
both field CVs were corrected for differential reddening with respect
to AKO~9 and scaled to the distance of 47~Tuc. For BV Cen, we took
$E_{B-V} = 0.15$ and $d = 500$~pc from Williger et al. (1988); for GK
Per, we took $E_{B-V} = 0.3$ and $d = 470$~pc from Wu et al. (1989). 
%For details of the IUE observations, see Williger et  al. (1988) for
%BV Cen and Wu et al. (1989) for GK Per.  

The overall similarity of the three spectra in Figure~11 is obvious
and supports the classification of AKO~9 as a CV. However, two
other points are also worth noting. First, the spectrum of AKO~9 is
both brighter and bluer than those of the two field CVs. Power law
fits to the latter, using 
the continuum regions indicated in Figure~3, yielded $\alpha = 0.93 
\pm 0.47$ for GK Per and  $\alpha = -1.19 \pm 0.24$ for BV Cen. As
expected, both values are considerably larger than the estimate
$\alpha = -2.74 \pm 0.11$ we obtained for AKO~9 (Section~\ref{avespec}). GK~Per's
extremely red spectral shape and FUV faintness is actually a
well-known anomaly and has been ascribed to the fact that the system
is also an intermediate polar (Watson, King \& Osborne 1985) in which
the inner accretion disk is truncated by the magnetic field of the
WD primary (Wu et al. 1989; Yi \& Kenyon 1997). BV Cen is not
thought to be a magnetic system, but Williger et al. (1988) have
already pointed out that its  FUV spectrum is redder than expected 
for a standard, optically thick disk. The canonical spectrum of such a 
disk -- based on describing  it as an ensemble of blackbodies -- is 
$F_{\lambda} \propto \lambda^{-7/3}$ (e.g. Frank, King \& Raine
2002). Thus AKO~9's FUV spectrum actually comes closest to this
theoretical prediction. 

Second, the emission line spectrum of AKO~9 more closely resembles
that of GK~Per than that of BV Cen. In particular, the strength of the
He~{\sc ii} line relative to C~{\sc iv} is quite unusual for a
CV. This could be partly due to the lower metallicity of this system,
although the evidence for saturation in C~{\sc iv}
(Section~\ref{slow}) would suggest otherwise. 
Strong optical (4686~\AA) or ultraviolet (1640~\AA) He~{\sc ii}
emission is sometimes used to classify systems as intermediate polars,
since it requires a strong extreme ultraviolet (EUV) source, such as the
accreting pole of a magnetic WD (e.g. Grindlay et
al. 1995). However, AKO~9's high FUV luminosity implies a
correspondingly high accretion rate (see Section~\ref{mdot}). It may 
therefore harbour a luminous, EUV-bright boundary layer at the
interface between the disk and the WD surface (e.g. Popham \& Narayan
1995). The strong He~{\sc ii} emission could thus be a signature of
AKO~9's high accretion rate, rather than of a magnetically channelled
accretion flow.

\subsection{Compatibility between Accretion Rate and Dwarf Nova
Behaviour}
\label{mdot}

The theoretical interpretation of dwarf nova outbursts require these
systems to accrete below a critical rate $\dot{M}_{crit}$. Accretion
above this rate quenches the dwarf nova instability and is 
characteristic of the steadily accreting nova-like variables
(e.g. Warner 1995). Given AKO~9's high FUV luminosity, it is clearly
worth asking if its accretion rate is, in fact, below the critical
rate for dwarf nova behaviour. 

In order to answer this question, we first convert the FUV luminosity
estimates from Section~\ref{lum} into constraints on the quiescent
accretion rate, $\dot{M}_{q}$. The total accretion luminosity
available is $L_{q} = 
G M_{WD} \dot{M}_{q} / R_{WD}$, although only half of this would be
expected to be liberated in the accretion disk, with the other half
being radiated by the boundary layer. A firm lower limit on
$\dot{M}_{q}$~is therefore given by $\dot{M}_{q} =
(L_{iso(1500-1800)} R_{WD}) / (G M_{WD})$. This yields $\dot{M}_{q} >
9.4 \times 10^{15}$~g~s$^{-1}$~for a 1~$M_{\odot}$~Hamada-Salpeter (1961)
WD and $\dot{M}_{q} > 3.2 \times 10^{16}$~g~s$^{-1}$~for a
0.5~$M_{\odot}$~WD. We can also obtain an 
approximate upper limit on $\dot{M}_{q}$~by replacing 
$L_{iso(1500-1800)}$~with $L_{disk(1000-2000)}$~and assuming that this 
represents half of the total accretion luminosity. In converting
$L_{iso}$~to $L_{disk}$, we then also use the maximum correction factor
allowed by our inclination constraints (Section~\ref{lum}). This
yields $\dot{M}_{q} < 3.6 \times 10^{17}$~g~s$^{-1}$~for a
1~$M_{\odot}$~WD and $\dot{M}_{q} < 1.2 \times
10^{18}$~g~s$^{-1}$~for a a 0.5~$M_{\odot}$~WD. We combine these
numbers to arrive at global constraints of $10^{16}~{\rm g~s^{-1}} \ltappeq \dot{M}_{q}
\ltappeq 10^{18}{\rm g~s^{-1}}$.

Strictly speaking, this $\dot{M}_{q}$~estimate cannot be compared
directly to $\dot{M}_{crit}$, since the former reflects the {\em transfer
rate through the disk in quiescence}, whereas the latter is a
constraint on $\dot{M}_2$, the {\em average mass transfer rate from the
secondary}. In general, the quiescent accretion rate may or may not reflect
$\dot{M}_2$, depending on the dwarf nova duty cycle and on the ratio
of accretion rates in outburst and quiescence. In the case of GK~Per,
the accretion luminosity during 
outbursts exceeds that during quiescence by roughly an order of
magnitude (Wu et al. 1989), but the average duration of quiescent
periods exceeds that of eruptions by a similar factor. Thus while the
quiescent accretion rate is, in principle, only a lower limit on $\dot{M}_2$, 
the example of GK~Per suggests that the two quantities are unlikely to 
differ by a large factor.

Taking $\dot{M}_{q}$ as a proxy for $\dot{M}_2$ in AKO~9, we find
that our observational constraints are below the theoretically
determined $\dot{M}_{crit}$. More specifically, Warner (1995) 
has provided a convenient reformulation of the critical
accretion rate calculated by Faulkner, Lin \& Papaloizou (1983),
\begin{equation}
\dot{M}_{crit} = 8.08 \times 10^{15}
\left(\frac{\alpha_H}{0.3}\right)^{3/10} \left(1 + q\right)^{7/8}
P_{orb(hr)}^{7/4} {\rm ~g~s^{-1}}
\end{equation}
(Warner 1995; Eq. 3.18b). 
In this equation, $\alpha_H$~is the Shakura-Sunyaev viscosity
parameter, and $P_{orb(hr)}$~is the orbital period, measured in
hours. Neglecting the weak dependence on $\alpha_H$, we therefore have
$\dot{M}_{crit} > 2.5 \times 10^{18}$~g~s$^{-1}$ for $P_{orb} =
1.109~d$, irrespective of mass ratio. Thus AKO~9's luminosity and accretion
rate are not in conflict with the occurrence of dwarf nova outbursts
in this system.

\subsection{The ``Unusual Brightening'' Revisited}
\label{brightening}

There is one other possible objection to the classification of AKO~9
as a dwarf nova. Minniti et al. (1997) inferred a roughly 1~hr rise
time for the low-state-to-high-state transition from  a rapid
brightening of the system in their U-band HST observations. This would
seem to argue against a DN interpretation, since the rise times of DN
eruptions are generally much longer. 

However, we already suggested in Knigge et al. (2002) that the
brightening observed by Minniti et al. may not have been a true
low-state-to-high-state transition, but instead an egress from eclipse
at a time when AKO~9 was already in a high state.  Minniti et
al. apparently discarded 
this possibility because the brightness increase they observed was in
excess of 2~mags (in U), and hence considerably larger than the 1~mag
eclipse depth seen in the low state. However, dwarf nova high states are
caused by increases in the instantaneous accretion luminosity, so the
fractional contribution of the accreting primary to the total light
will be much larger in the high state than in the low state. The primary
eclipse should therefore be expected to be much deeper in the high
state.

The new ephemerides we derived in Section~\ref{ephem} allow us to test
this idea quantitatively. The observations described by Minniti et
al. (1997) took place on 1995 
October 25, and consist of fifteen 350-s exposures, with the first
starting at 16:08 UT and the last ending at 20:32 UT. Thus their
observations cover 16\%~of AKO~9's orbital cycle. If the brightening
they observed is, in fact, an 
eclipse egress, we can estimate the corresponding time of mid-eclipse 
from the half depth point of their brightening event. From their 
light curve (Figure~2 in Minniti et al. 1997), we estimate this point
to be about half-way between their 5th and 6th exposures. Now the {\em
full} eclipse width at half depth is $\Delta_{\phi} = 0.090$
(Section~\ref{qi}), so this half depth timing must be shifted by
$ - P_{orb} \times \Delta_{\phi} / 2$ to convert it to the
corresponding time of mid-eclipse. The timing thus obtained
corresponds to about 16:20 UT, i.e. between  their first and second
exposures. Given that this new timing is closest to the 1993 eclipse,
we then calculate the predicted orbital phase from
Equation~\ref{eq_ephem2} (rather than Equation~\ref{eq_ephem1}, which 
does not include the 1993 timing). Adopting a conservative error
estimate of $\pm 10$~minutes for the 1995 timing yields a predicted 
orbital phase of $0.998 \pm 0.019$. Thus the brightening observed by 
Minniti et al. was indeed an eclipse egress.

Before concluding this section, we note that Equation~\ref{eq_ephem1}
predicts an orbital phase of $0.965 \pm 0.011$~for the 1995
timing. Thus, as already noted in Section~\ref{ephem}, both the 1993
and 1995 timings are not quite consistent with the linear ephemeris
derived from the most recent eclipse timings.

\subsection{Formation and Evolution of AKO~9: Past, Present and
Future}
\label{evo}

AKO~9 is located close to the centre of 47~Tuc and well within the
core radius of this cluster (e.g. Paper I). This alone makes it
extremely unlikely that the system descended directly (without
intervening dynamical encounters) from a primordial binary system. As 
shown by Davies (1997), such potential CV progenitors simply cannot 
survive in the hostile environment provided by a dense cluster
core. ``Primordial CVs'' may exist in globulars, but only in the
outskirts of clusters with long relaxations times. Thus AKO~9 is
almost certainly an example of a CV formed by dynamical means, either
via tidal capture or in a three-body encounter.

Given that AKO~9's secondary appears to be a sub-giant, it seems
likely that mass transfer in this system started relatively recently,
as a direct consequence of the nuclear evolution of the
secondary. More specifically, mass transfer was probably initiated
when the radius of the donor star caught up with the Roche lobe during
the donor's evolution from the MS to the RGB via the Hertzsprung
gap. One can check this idea by considering the time-scale for radius
evolution of the secondary. We have obtained a crude estimate of this 
time scale by comparing the radii of 0.9 $M_{\odot}$~stars in the
10~Gyr and 12~Gyr isochrones of VandenBerg (2000). A star of this mass
is located near the beginning (end) of the sub-giant branch in the 
10~Gyr (12~Gyr) isochrone and has a radius of about
1.2~$R_{\odot}$~(2.3~$R_{\odot}$). The time-scale for radius expansion
due to nuclear evolution is therefore roughly $\tau_{nuc} = R/\dot{R}
\simeq 3$~Gyr. This number should probably be viewed as a lower limit,
since nuclear evolution accelerates across the Hertzsprung gap and
AKO~9's secondary is located near the end of the sub-giant branch. 
If nuclear expansion is responsible for driving the mass transfer, 
we expect $\tau_{nuc}$~to be be close to the mass-transfer time scale,
$\tau_{MT} = | M_2/\dot{M_2}|$.  
In Section~\ref{mdot}, we estimated that $\dot{M}_2 \sim \dot{M}_{q}
\simeq 10^{17 \pm 1}{\rm ~g~s^{-1}}$. Combining this with our estimate
for the mass of 
the secondary $M_2 \simeq 0.9~M_{\odot}$, we find that $\tau_{MT}
\simeq 6 \times 10^{8 \pm 1}$~yrs. Thus the mass-transfer and
nuclear-evolution time scales are indeed compatible.

We finally consider the possible endpoints of AKO~9's evolution as a
CV. Stable, nuclear time-scale mass transfer will likely continue as
the donor star expands on the sub-giant branch and ultimately moves up
the giant branch. This evolution will terminate either when the donor 
star has lost its entire envelope or once the accreting
WD reaches the Chandrasekhar limit. In the first case, AKO~9 will
become a detached binary, with a low-mass Helium WD secondary in orbit
around a more massive WD primary. The second case requires that the WD is 
able to hang on to the mass it accretes from the secondary (i.e. the
mass transfer must be roughly conservative). This contrasts with
ordinary CVs, which are thought to expel most of their accreted
material during nova explosions. However,  King, Rolfe \& Schenker
(2003) have recently shown that long-period dwarf novae (like AKO~9) 
may be able to avoid this, making them plausible Type Ia supernova
progenitors. If the WD in AKO~9 does undergo accretion-induced
collapse in the future, the post-SN fate of the system may be as a
low-mass X-ray binary, with a Helium WD secondary orbiting a neutron
star primary. At least one such binary pulsar system is already known
in 47~Tuc (Edmonds et al. 2001), but note that accretion-induced
collapse is not the only formation mechanism for such systems
(Verbunt, Lewin \& van Paradijs 1989).

%The key point in their argument is that the mass
%accretion rates during the outbursts of these system may be high
%enough to allow steady nuclear burning, thus precluding the explosive
%burning of the same material in a nova eruption. 

\section{Conclusions}
\label{conclusions}

We have presented and analysed time-resolved, far-ultraviolet (FUV)
spectroscopy and photometry of the 1.1~day eclipsing binary system
AKO~9 in 47 Tuc. AKO~9's FUV spectrum is blue, with prominent C~{\sc iv} and
H~{\sc ii} emission lines, and broadly resembles that of long-period,
cataclysmic variables in the galactic field. By combining our
time-resolved FUV data with archival optical photometry, we have been
able to derive a precise ephemeris for this system and to constrain
several of its key parameters. 

All of the observational evidence we have gathered is consistent with 
AKO~9 being a long-period, dwarf-nova-type CV, in 
which mass  transfer is driven by the nuclear expansion of a sub-giant donor
star. In particular, we showed that the ``unusual brightening'' 
described by Minniti et al. (1997) was simply an eclipse egress
observed at a time when AKO~9 was near the peak of a dwarf nova
eruption. We therefore conclude that AKO~9 is the first
spectroscopically confirmed cataclysmic variable in 47~Tuc. 

We have also considered the likely formation and evolution of AKO~9. 
We found that the system's location near the cluster center suggests a
dynamical formation mechanism, i.e. tidal capture or a 3-body
encounter. AKO~9's CV phase will come to an end when the secondary has 
lost its entire envelope or the accreting WD reaches the Chandrasekhar
limit. In the former case, the system will become a double WD 
system, with a low-mass Helium WD secondary in orbit around a more
massive WD primary. In the latter case, the accreting WD will undergo
accretion-induced collapse and explode in a Type Ia supernova.

\acknowledgments We are grateful to the anonymous referee for a prompt 
and helpful report. Support for proposal \#8219 was provided by NASA
through a grant from the Space Telescope Science Institute, which is
operated by the Association of Universities for Research in Astronomy,
Inc., under NASA contract NAS 5-26555.

\clearpage

\begin{deluxetable}{lccc}
\tablewidth{390pt}
\footnotesize
\tablecaption{Eclipse timings for AKO~9}
\tablehead{
\colhead{Month/Year}&
\colhead{Bandpass}&
\colhead{~~~Cycle Count\tablenotemark{a}~~~}&
\colhead{Time of Mid-Eclipse\tablenotemark{b}}}
\startdata
9/1993 & U band (F336W) & -2083 & 9236.4623(20)\\
7/1999 & V band (F555W) & -162  & 11367.10320(66)\\
9/1999 & FUV            & -103  & 11432.5374(12)\\
8/2000 & FUV            &  199  & 11767.48749(67)\\
\enddata
\tablenotetext{a}{Cycle numbers are relative to the ephemeris defined
by Equation~\ref{eq_ephem1}}
\tablenotetext{b}{Given as BJD-2440000; the numbers in parentheses
correspond to the formal errors on the last two digits.} 
\end{deluxetable} 

\clearpage

\begin{deluxetable}{lll}
\tablewidth{480pt}
\footnotesize
\tablecaption{Inferred System Parameters}
\tablehead{
\colhead{Parameter\tablenotemark{a}}&
\colhead{Value}&
\colhead{Comment}}
\startdata
$P_{orb}$     & $1.1090989 \pm 0.0000074$  &{\footnotesize from eclipse timings (Section~\ref{ephem})} \\
$M_2$         & $0.9~M_{\odot}$            &{\footnotesize from CMD location and density constraint (Section~\ref{secpars})} \\
$R_2$         & $2.1~R_{\odot}$            & {\footnotesize from CMD location and density constraint (Section~\ref{secpars})} \\
$T_{eff,2}$   & $4900$~K                   &{\footnotesize from CMD location and density constraint (Section~\ref{secpars})} \\
$\log{g}$     & $3.8$                      & {\footnotesize from CMD location and density constraint (Section~\ref{secpars})} \\
$q = M_2/M_1$ & $0.6 - 2.6$               & {\footnotesize from eclipse width and ellipsoidal amplitude (Section~\ref{qi})\tablenotemark{b}} \\
$i$           & $68^o - 80^o$             & {\footnotesize from eclipse width and ellipsoidal amplitude (Section~\ref{qi})\tablenotemark{b}} \\
$M_1$         & $0.3~M_{\odot}$~-~$1.3~M_{\odot}$ & {\footnotesize from constraints on $q$~and $M_2$} \\
$\dot{M}_{q}$ & $10^{17 \pm 1}{\rm ~g~s^{-1}}$ & {\footnotesize quiescent accretion rate; from FUV luminosity (Section~\ref{mdot})}\\
\enddata 
\tablenotetext{a}{Parameters subscripted with ``1'' and ``2'' refer to
the primary and secondary stars, respectively.}
\tablenotetext{b}{The constraints on the parameters $q$~and $i$~are
strongly correlated; see Section~\ref{qi}~and Figure~10.}
\end{deluxetable}

\clearpage

%% No more than seven \figcaption commands are allowed per page,
%% so if you have more than seven captions, insert a \clearpage

%% There must be a \figcaption command for each legend. Key the text of the
%% legend and the optional \label in curly braces. If you wish, you may
%% include the name of the corresponding figure file in square brackets.
%% The label is for identification purposes only. It will not insert the
%% figures themselves into the document.
%% If you want to include your art in the paper, use \plotone.
%% Refer to the on-line documentation for details.

\figcaption[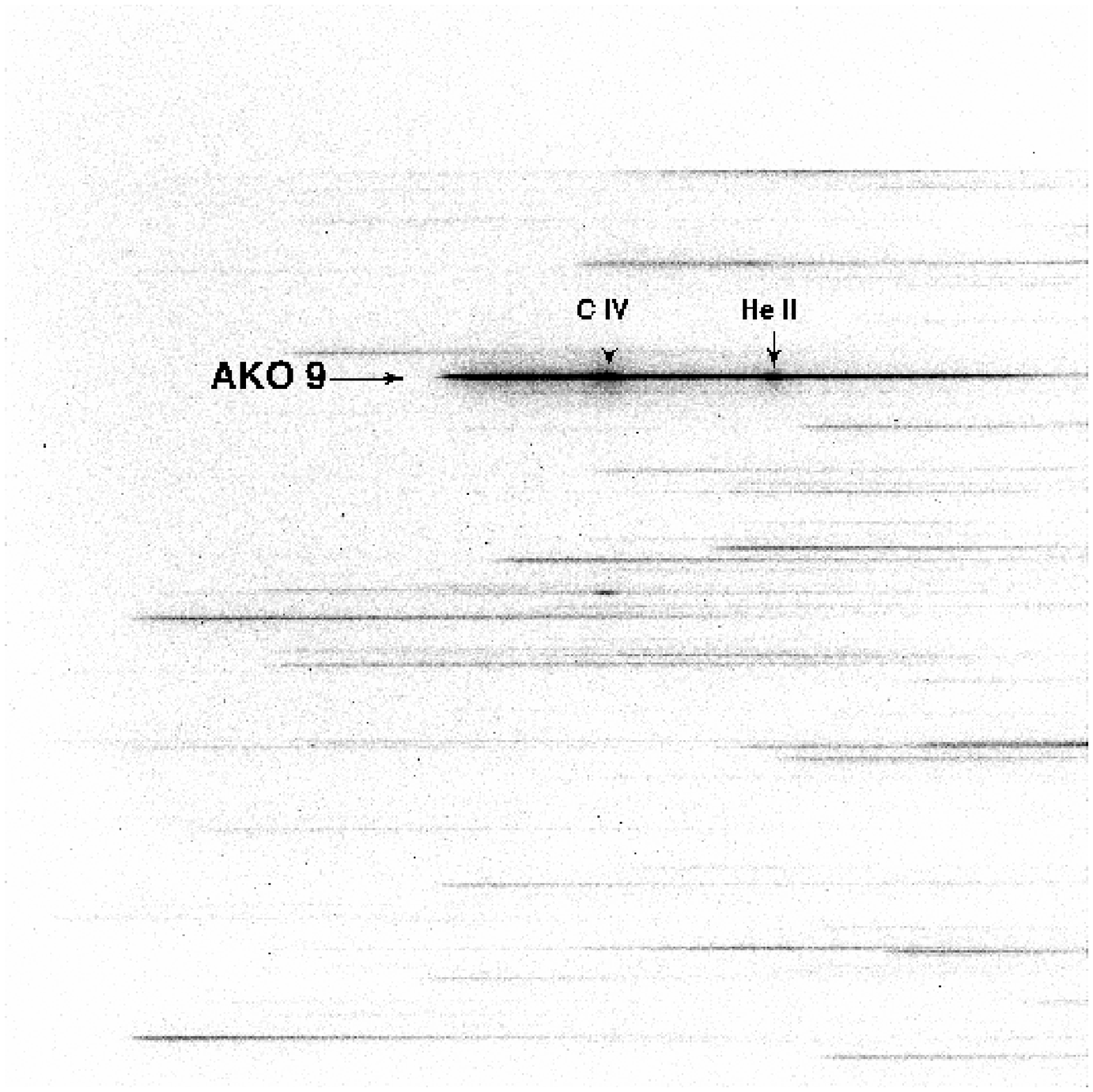]{The summed 2-D spectral image obtained during
Epoch~3. Each trail in this figure corresponds the dispersed image of
a FUV point source. The sharp cut-off at the left hand side of each
trail is due to the abrupt decrease in sensitivity around 1450~\AA,
where the quartz filter becomes opaque. The spectrum of AKO~9 is 
marked, as are the locations of the two strong emission lines in this
spectrum.}

\figcaption[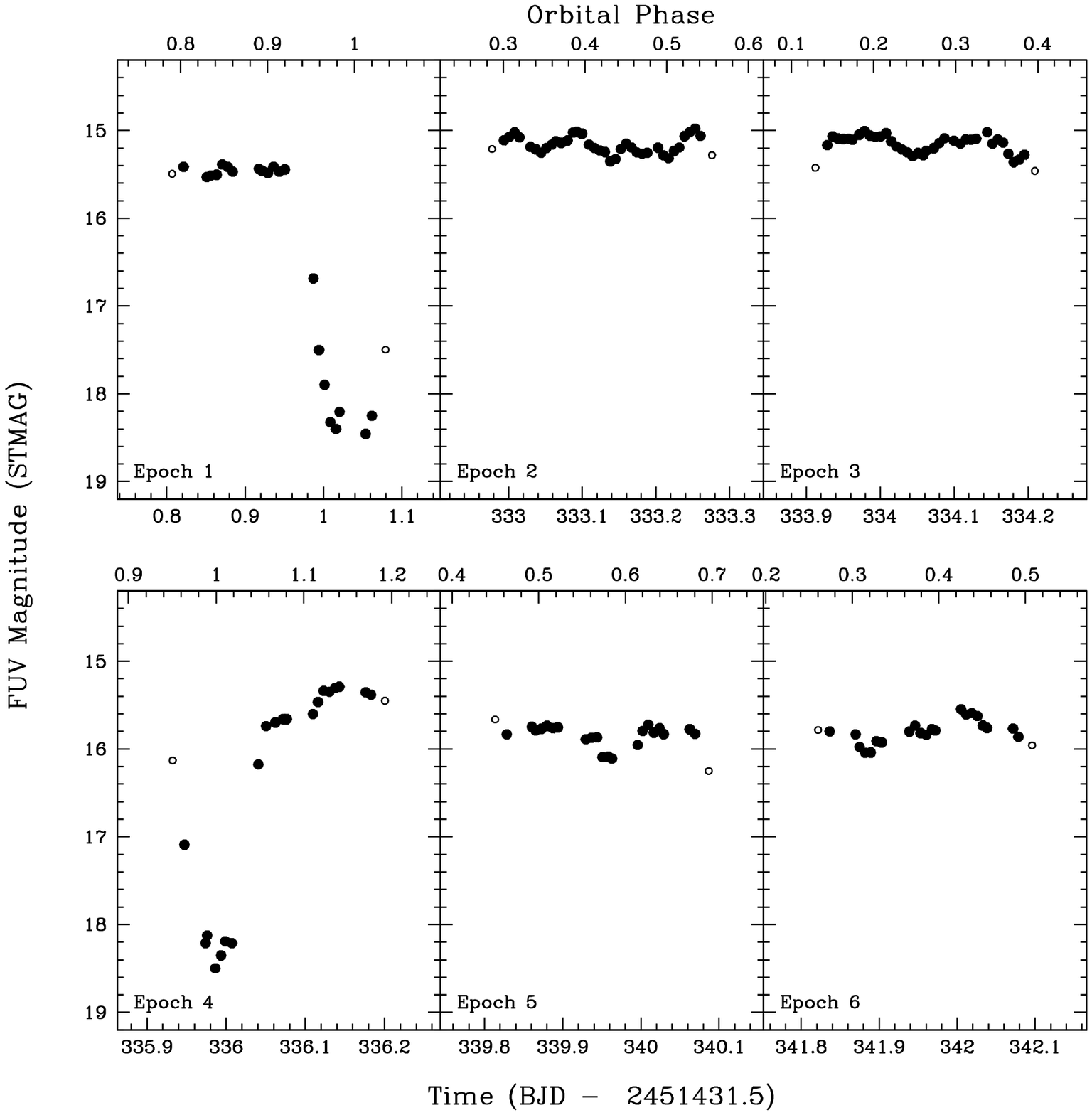]{The FUV light curve of AKO~9 over the course of
our six observing epochs. Filled circles correspond to magnitudes estimated
from the FUV spectroscopy; open circles mark magnitudes derived from
imaging. Magnitudes are expressed on the STMAG system. Orbital
phases corresponding to the times of observation have been calculated
from Equation~\ref{eq_ephem1}~and are indicated at the top of each
panel.}

\figcaption[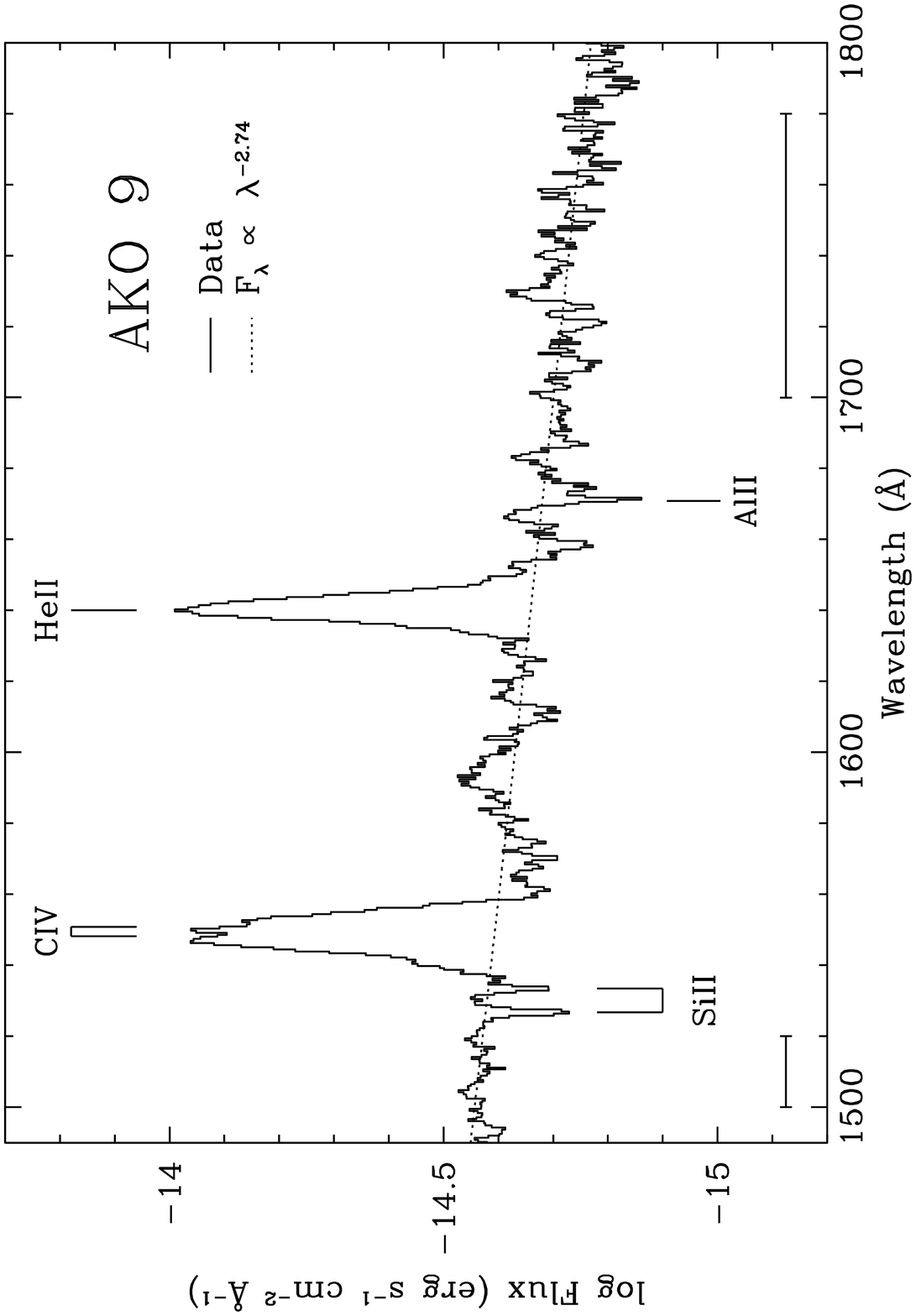]{The average out-of-eclipse FUV spectrum of
AKO~9. This spectrum is an exposure-time-weighted average of all FUV
spectra from all observing epochs, excluding only points affected by 
the primary eclipses. The rest wavelengths of the C~{\sc iv}~doublet
and He~{\sc ii}~are marked, as are those of the narrow absorption lines  
due to Si~{\sc ii}~and Al~{\sc ii}. The Si~{\sc ii}~lines were used to 
wavelength calibrate the spectrum. The Al~{\sc ii} line is probably 
interstellar. The dotted line shows a power law fit to the continuum
windows indicated by the horizontal bars near the bottom of the
plot. The observed spectrum was corrected for reddening before
carrying out the fit, but the spectra shown here are the uncorrected
data and the reddened power law model. The best-fitting power law
index was $\alpha=-2.74 \pm 0.11$.}

\figcaption[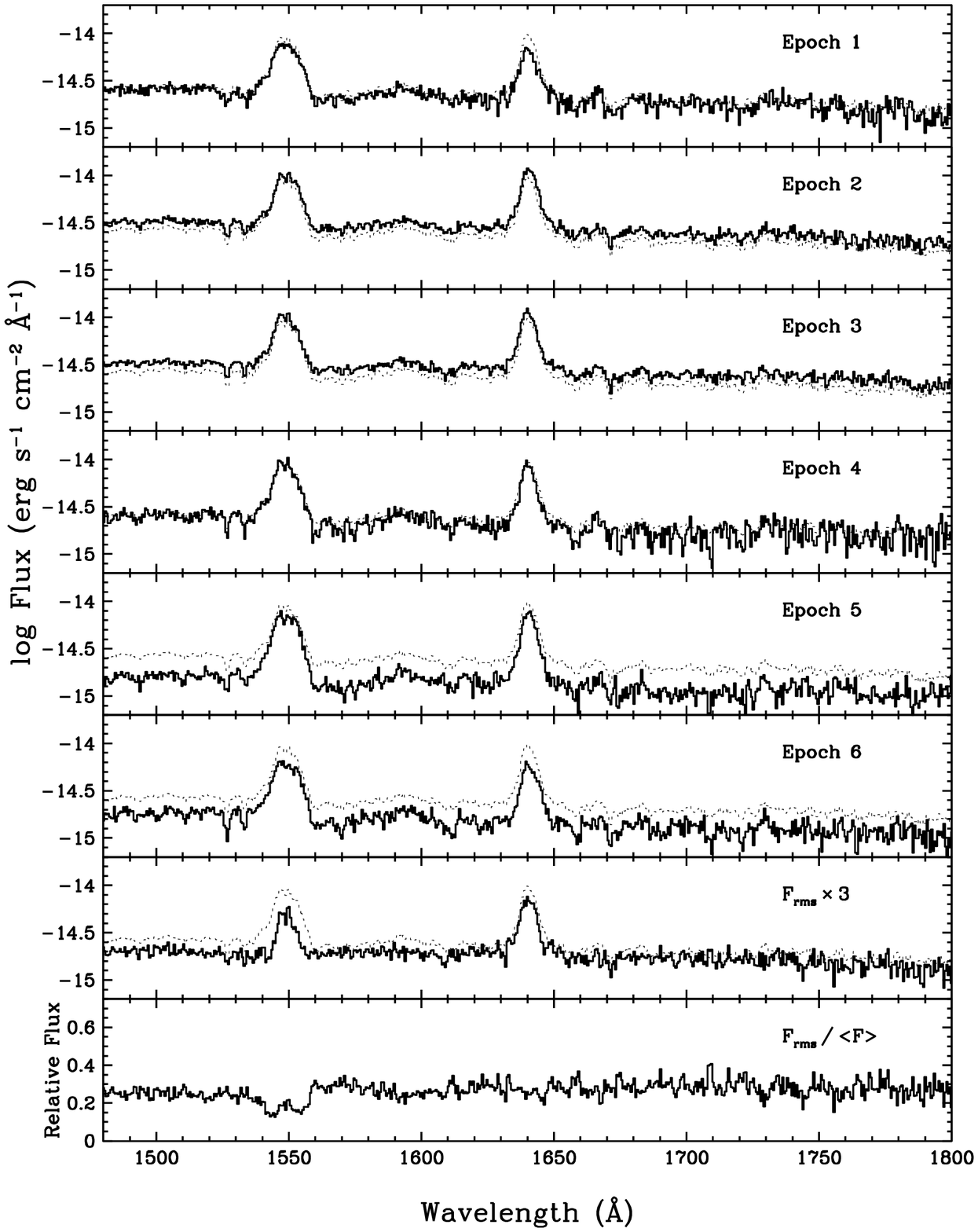]{The average FUV spectrum in each of the six
observing epochs. Spectra affected by primary eclipses were again
excluded in constructing the averages. The dotted line in each panel
shows the average spectrum across all observing epochs for comparison
(cf. Figure~3). The bottom two panels shows the corresponding rms
spectrum (cf. Equation~\ref{rms}), first in units of $F_{\lambda}$
(after multiplying the rms spectrum by a factor 3), and then relative
to the global mean spectrum.}

\figcaption[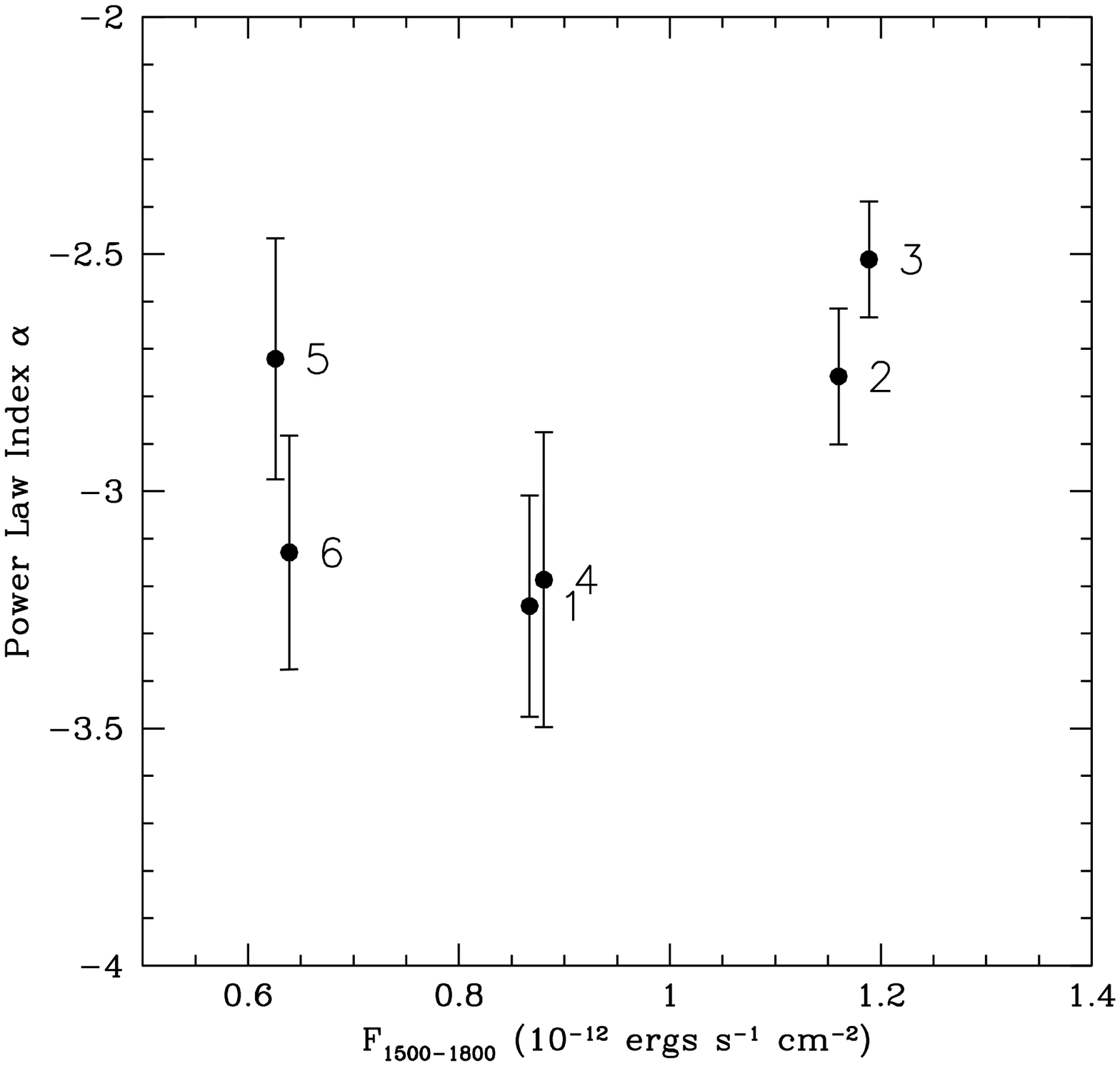]{Spectral index estimates versus integrated FUV
flux as estimated from the average spectra of the six observing
epochs. The spectral indices were obtained from power law fits of the
form $F_\lambda \propto \lambda^\alpha$~to the de-reddened average
spectra. The integrated flux was calculated over the interval
1500~--~1800~\AA.}

\figcaption[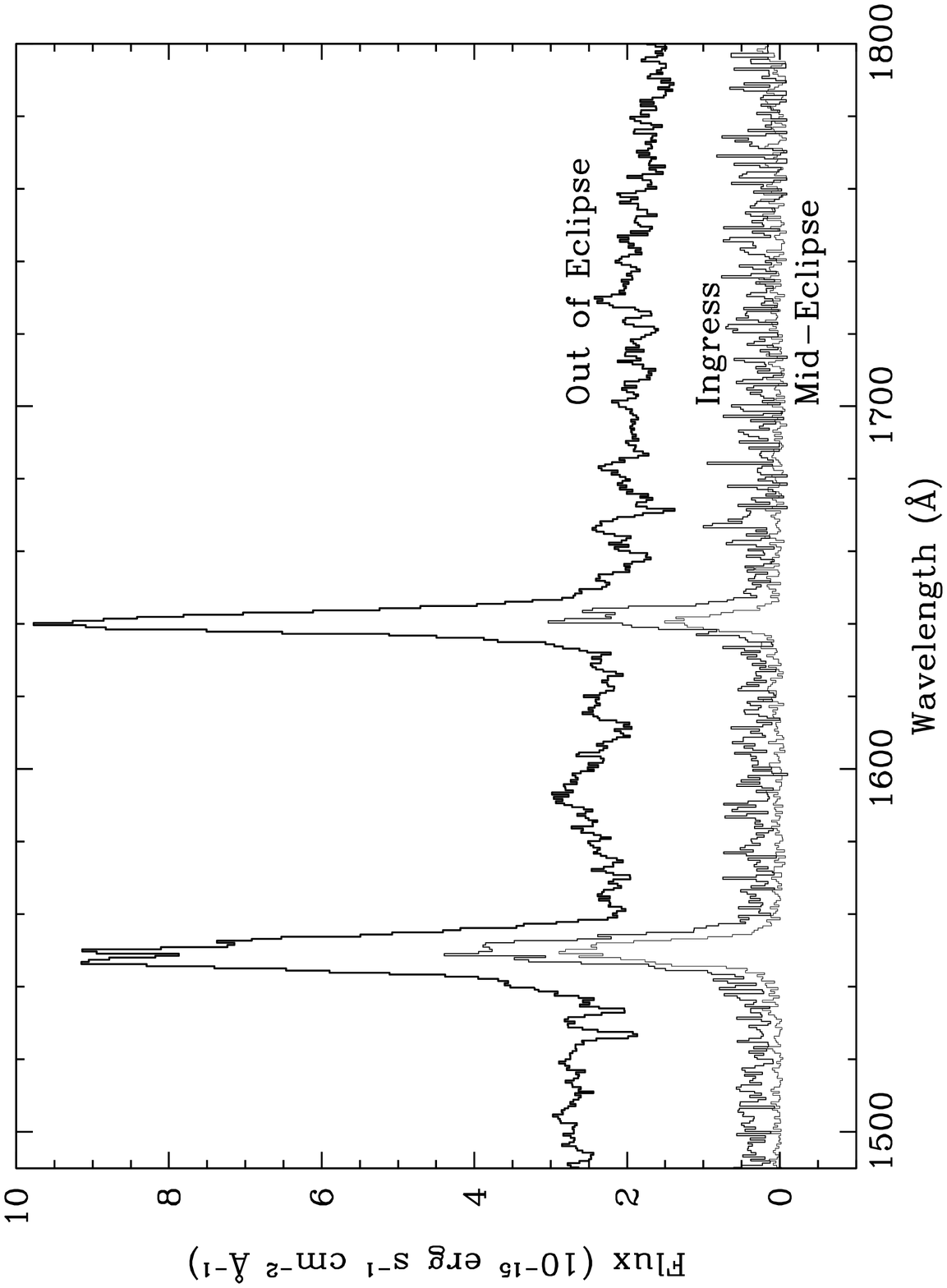]{The average out-of-eclipse, ingress, and
mid-eclipse spectra derived from the FUV data. The out-of-eclipse
spectrum is the same as that shown in Figure~3. The definition of the 
ingress and mid-eclipse spectra is given in
Section~\ref{phasespec}. Note that the strong emission  lines are much
less deeply eclipsed than the FUV continuum.}

\clearpage 

\figcaption[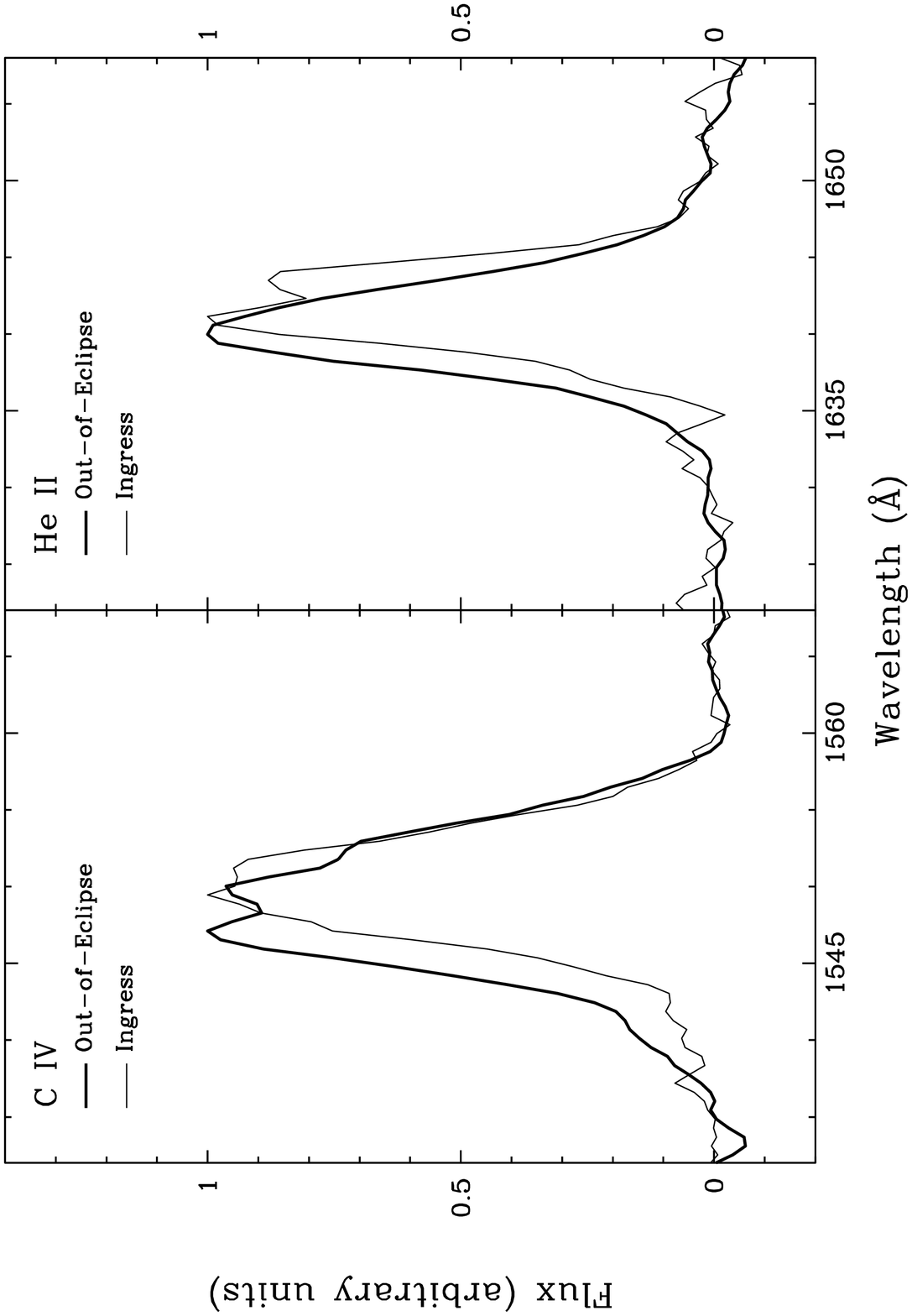]{A comparison of the out-of-eclipse and ingress
line profile shapes for C~{\sc iv}~and He~{\sc ii}. For both lines, the
thick line corresponds to the out-of-eclipse profile and the thin line
to that derived from the average ingress spectrum. All profiles have
been continuum subtracted, smoothed with a 3-point boxcar filter and
rescaled to a peak flux of unity for ease of comparison. Note that the
blue line wing of the ingress profile is suppressed relative to the
red wing, a clear signature of a rotational disturbance.}

\figcaption[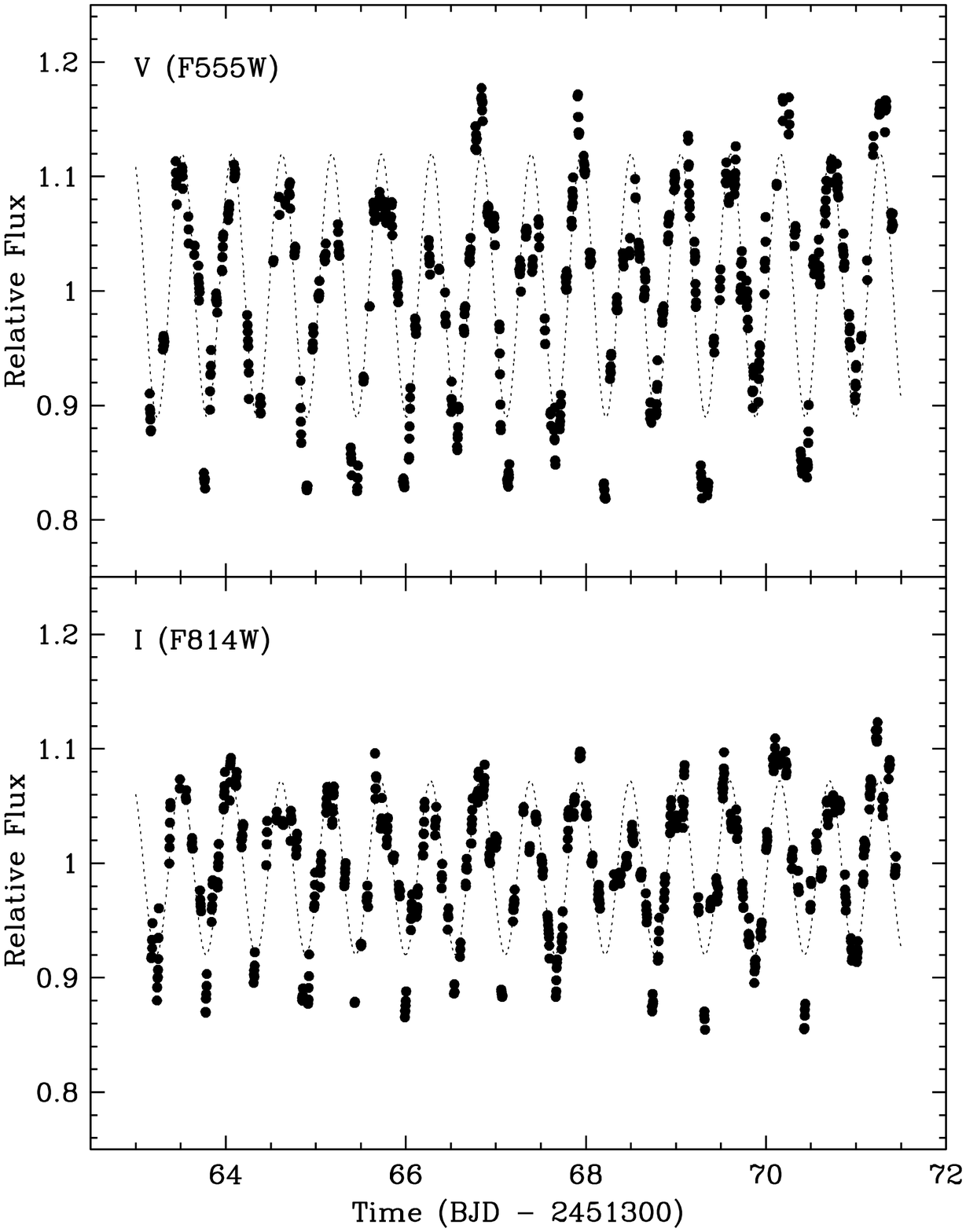]{The V- and I-band light curves of AKO~9 obtained
from the Gilliland et al. (2000) observations. A strong signal on roughly 0.55 days
(half the orbital period) is easily seen in both light curves. The
dotted lines are the best-fitting sine waves for the joint period
estimate of 0.554~days.}

\figcaption[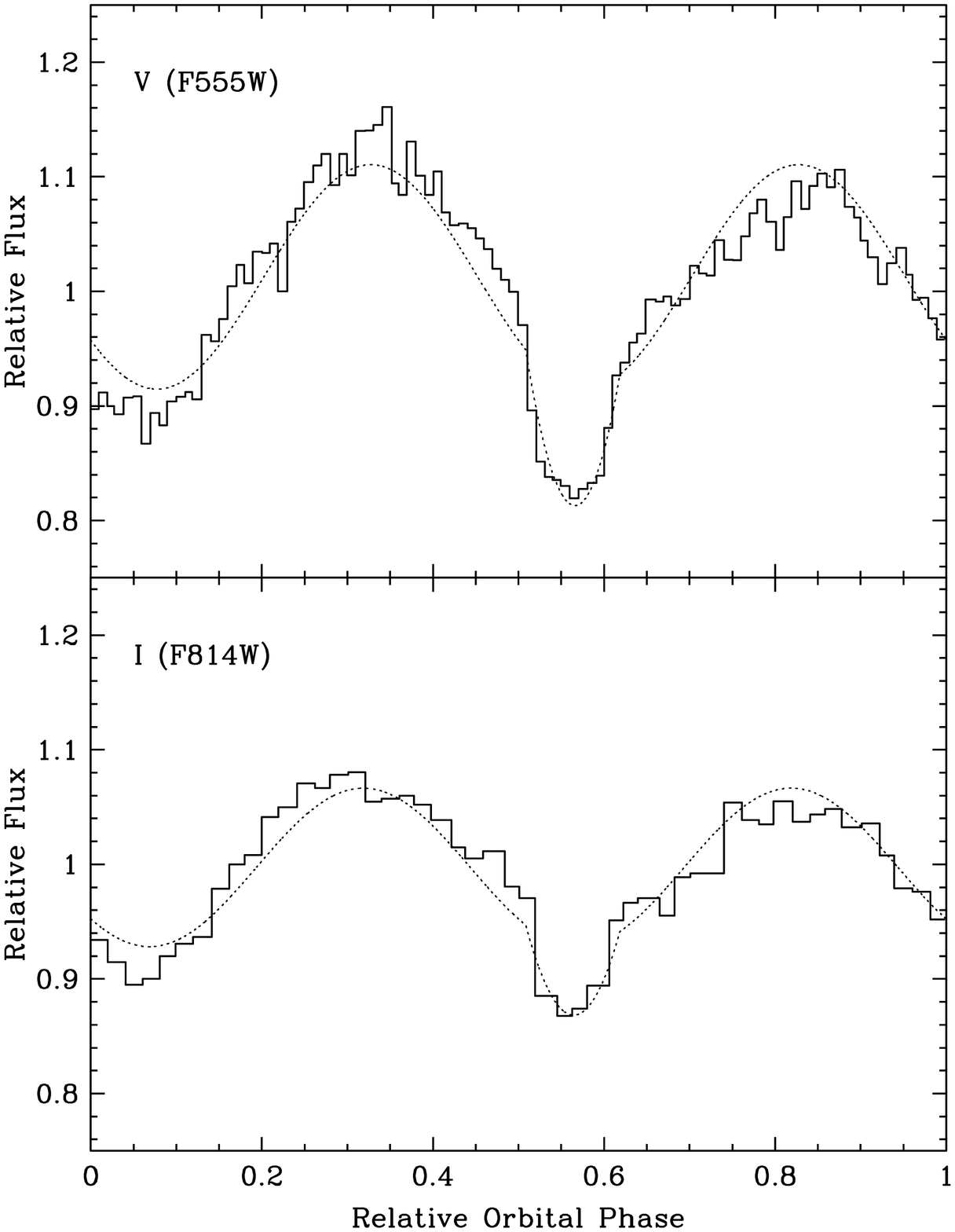]{The optical data folded onto a trial period of $P
= 1.108$~days. The absolute phasing is arbitrary, with the first V-band point
having been assigned phase 0.0. The V-band observations have been 
binned into 100 phase bins. The I-band observations do not cover the
eclipse as well, so a coarser phase grid of 50 bins was used for these
data points. A double-humped (ellipsoidal) modulation is clearly 
present in both V and I, but a weak primary eclipse is also visible in
both filters. The dotted lines show the simple, 3-component model
fit to the data described in the text.} 

\figcaption[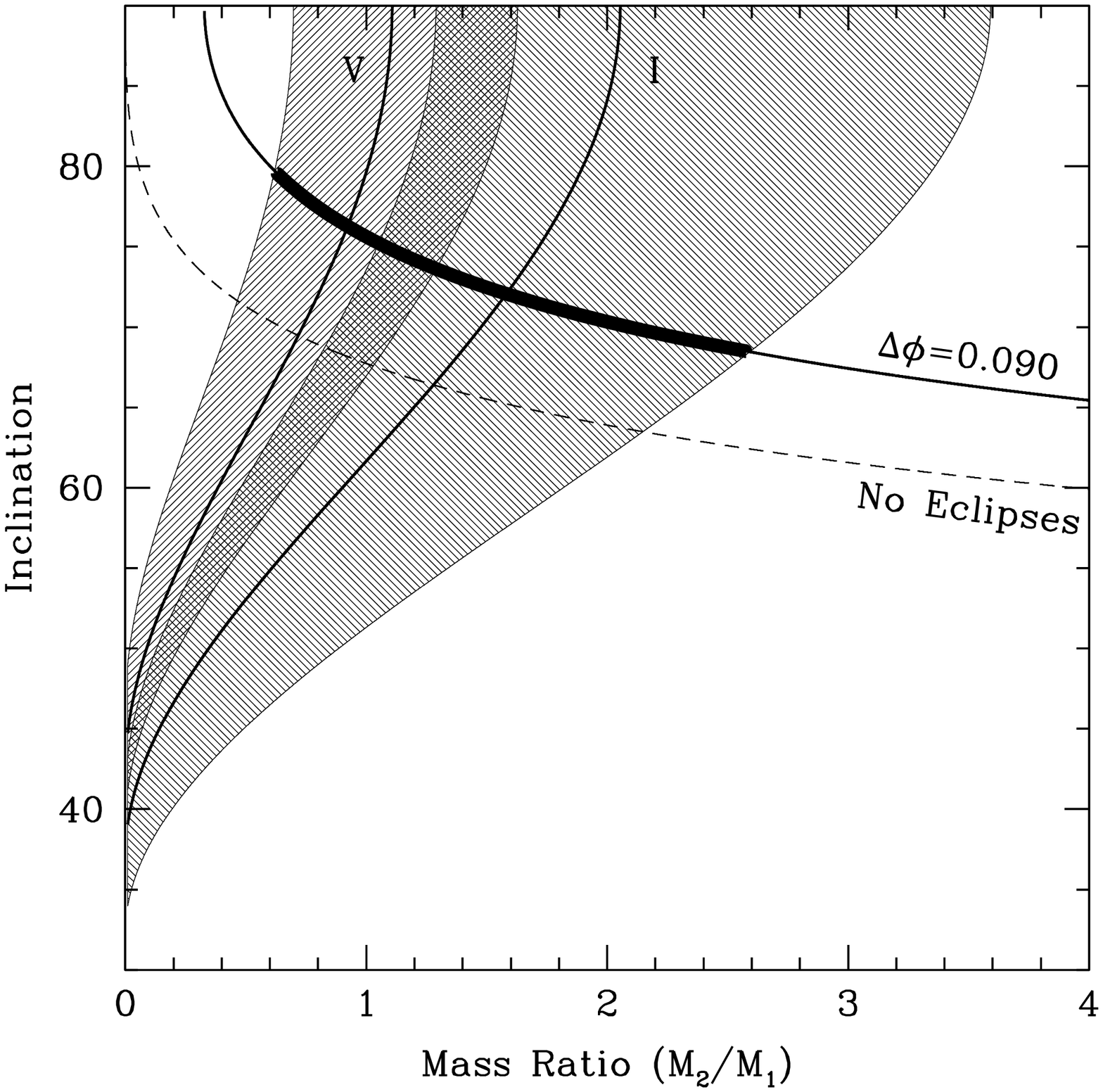]{Mass ratio and inclination constraints on
AKO~9. The line labelled ``$\Delta_{\phi} = 0.090$'' shows the family of
solutions in the ($q,i$)-plane that produce this eclipse width. The
error on this constraint resulting from the error on
$\Delta_{\phi}$~($\pm 0.002$) is negligible. The dashed line labelled
``No Eclipses'' shows the strict lower limit on the inclination
imposed by the sheer existence of eclipses in this system. The two
lines labelled ``V'' and ``I'' mark the families of ($q,i$)-solutions
derived from the measured ellipsoidal amplitudes in the two optical
bandpasses. The light-shaded regions surrounding these lines indicate
the error on these constraints, and the dark-shaded region marks the
area where these errors overlap. The thick line segment on the
``$\Delta_{\phi} = 0.090$'' curve shows the set of ($q,i$)-solutions
that satisfies both eclipse and ellipsoidal constraints.}

\figcaption[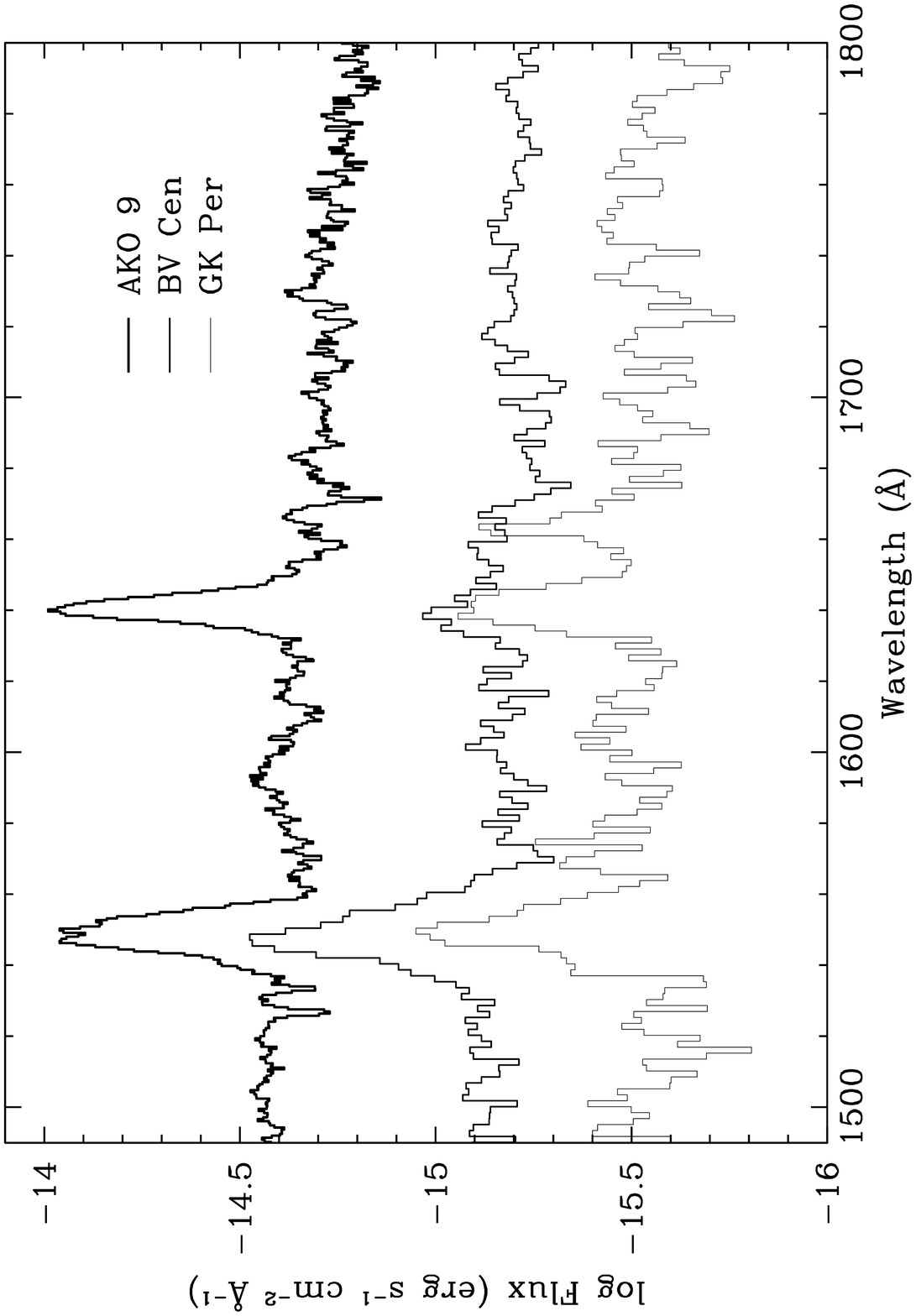]{A comparison of the average, out-of-eclipse FUV
spectrum of AKO~9 against the FUV spectra of two long-period field
CVs (BV~Cen: $P_{orb} = 0.61$~d; GK~Per: $P_{orb} = 2.00$~d). The
field CV spectra have been corrected for differential reddening
with respect to AKO~9 and scaled to the distance of 47~Tuc; see text
for details.}

\clearpage
\newpage
\pagebreak

\begin{figure}
\epsscale{1.0}
\figurenum{1}
\plotone{f1.eps}
\caption{}
\end{figure}

\clearpage
\newpage
\pagebreak

\begin{figure}
\epsscale{1.0}
\figurenum{2}
\plotone{f2.eps}
\caption{}
\end{figure}

\clearpage
\newpage
\pagebreak

\begin{figure}
\epsscale{1.0}
\figurenum{3}
\plotone{f3.eps}
\caption{}
\end{figure}

\clearpage
\newpage
\pagebreak

\begin{figure}
\epsscale{1.0}
\figurenum{4}
\plotone{f4.eps}
\caption{}
\end{figure}

\clearpage
\newpage
\pagebreak

\begin{figure}
\epsscale{1.0}
\figurenum{5}
\plotone{f5.eps}
\caption{}
\end{figure}

\clearpage
\newpage
\pagebreak

\begin{figure}
\epsscale{1.0}
\figurenum{6}
\plotone{f6.eps}
\caption{}
\end{figure}

\clearpage
\newpage
\pagebreak

\begin{figure}
\epsscale{1.0}
\figurenum{7}
\plotone{f7.eps}
\caption{}
\end{figure}

\clearpage
\newpage
\pagebreak

\begin{figure}
\epsscale{1.0}
\figurenum{8}
\plotone{f8.eps}
\caption{}
\end{figure}

\clearpage
\newpage
\pagebreak

\begin{figure}
\epsscale{1.0}
\figurenum{9}
\plotone{f9.eps}
\caption{}
\end{figure}

\clearpage
\newpage
\pagebreak

\begin{figure}
\epsscale{1.0}
\figurenum{10}
\plotone{f10.eps}
\caption{}
\end{figure}

\clearpage
\newpage
\pagebreak

\begin{figure}
\epsscale{1.0}
\figurenum{11}
\plotone{f11.eps}
\caption{}
\end{figure}

%% The following command ends your manuscript. LaTeX will ignore any text

\end{document}